\documentstyle[12pt,
epsfig]{article}
\begin{document}
\newcommand{\tcr}{T_{cr}}
\newcommand{\sxj}{\sigma_j}
\newcommand{\vt}{\tilde{v}}
\newcommand{\sxjt}{\tilde{\sigma_j}}
\newcommand{\dtt}{\tilde{\delta}}
\newcommand{\Lt}{\tilde{\Lambda}}
\newcommand{\Ut}{\tilde{U}}
\newcommand{\Ct}{\tilde{C}}
\newcommand{\phit}{\tilde{\phi}}
\newcommand{\rht}{\tilde{\rho}}
\newcommand{\done}{\delta\phi_1}
\newcommand{\dy}{\delta y}
\newcommand{\tr}{\rm Tr}
\newcommand{\sx}{\sigma}
\newcommand{\mpl}{m_{Pl}}
\newcommand{\Mpl}{M_{Pl}}
\newcommand{\lx}{\lambda}
\newcommand{\Lx}{\Lambda}
\newcommand{\kx}{\kappa}
\newcommand{\ex}{\epsilon}
\newcommand{\be}{\begin{equation}}
\newcommand{\ee}{\end{equation}}
\newcommand{\eesn}{\end{subequations}}
\newcommand{\besn}{\begin{subequations}}
\newcommand{\beq}{\begin{eqalignno}}
\newcommand{\eeq}{\end{eqalignno}}
\def \lta {\mathrel{\vcenter
     {\hbox{$<$}\nointerlineskip\hbox{$\sim$}}}}
\def \gta {\mathrel{\vcenter
     {\hbox{$>$}\nointerlineskip\hbox{$\sim$}}}}

\newcommand{\nwc}{\newcommand}
%
%
\nwc{\cl}  {\clubsuit}
\nwc{\di}  {\diamondsuit}
\nwc{\sps} {\spadesuit}
\nwc{\hyp} {\hyphenation}
\nwc{\ba}  {\begin{array}}
\nwc{\ea}  {\end{array}}
\nwc{\bdm} {\begin{displaymath}}
\nwc{\edm} {\end{displaymath}}
\nwc{\bea} {\be\ba{rcl}}
\nwc{\eea} {\ea\ee}
\nwc{\ben} {\begin{eqnarray}}
\nwc{\een} {\end{eqnarray}}
\nwc{\bda} {\bdm\ba{lcl}}
\nwc{\eda} {\ea\edm}
\nwc{\bc}  {\begin{center}}
\nwc{\ec}  {\end{center}}
\nwc{\ds}  {\displaystyle}
\nwc{\bmat}{\left(\ba}
\nwc{\emat}{\ea\right)}
\nwc{\non} {\nonumber}
\nwc{\bib} {\bibitem}
\nwc{\lra} {\longrightarrow}
\nwc{\Llra}{\Longleftrightarrow}
\nwc{\ra}  {\rightarrow}
\nwc{\Ra}  {\Rightarrow}
\nwc{\lmt} {\longmapsto}
\nwc{\pa} {\partial}
\nwc{\iy}  {\infty}
\nwc{\ovl}  {\overline}
\nwc{\hm}  {\hspace{3mm}}
\nwc{\lf}  {\left}
\nwc{\ri}  {\right}
\nwc{\lm}  {\limits}
\nwc{\lb}  {\lbrack}
\nwc{\rb}  {\rbrack}
\nwc{\ov}  {\over}
\nwc{\pr}  {\prime}
\nwc{\nnn} {\nonumber \vspace{.2cm} \\ }
\nwc{\Sc}  {{\cal S}}
\nwc{\Lc}  {{\cal L}}
\nwc{\Rc}  {{\cal R}}
\nwc{\Dc}  {{\cal D}}
\nwc{\Oc}  {{\cal O}}
\nwc{\Cc}  {{\cal C}}
\nwc{\Pc}  {{\cal P}}
\nwc{\Mc}  {{\cal M}}
\nwc{\Ec}  {{\cal E}}
\nwc{\Fc}  {{\cal F}}
\nwc{\Hc}  {{\cal H}}
\nwc{\Kc}  {{\cal K}}
\nwc{\Xc}  {{\cal X}}
\nwc{\Gc}  {{\cal G}}
\nwc{\Zc}  {{\cal Z}}
\nwc{\Nc}  {{\cal N}}
\nwc{\fca} {{\cal f}}
\nwc{\xc}  {{\cal x}}
\nwc{\Ac}  {{\cal A}}
\nwc{\Bc}  {{\cal B}}
\nwc{\Uc}  {{\cal U}}
\nwc{\Vc}  {{\cal V}}
%
%
\nwc{\Th} {\Theta}
\nwc{\th} {\theta}
\nwc{\vth} {\vartheta}
\nwc{\eps}{\epsilon}
\nwc{\si} {\sigma}
\nwc{\Gm} {\Gamma}
\nwc{\gm} {\gamma}
\nwc{\bt} {\beta}
\nwc{\La} {\Lambda}
\nwc{\la} {\lambda}
\nwc{\om} {\omega}
\nwc{\Om} {\Omega}
\nwc{\dt} {\delta}
\nwc{\Si} {\Sigma}
\nwc{\Dt} {\Delta}
\nwc{\al} {\alpha}
\nwc{\vph}{\varphi}
\nwc{\zt} {\zeta}
%
%
\def\tr{\mathop{\rm tr}}
\def\Tr{\mathop{\rm Tr}}
\def\Det{\mathop{\rm Det}}
\def\Im{\mathop{\rm Im}}
\def\Re{\mathop{\rm Re}}
\def\secder#1#2#3{{\partial^2 #1\over\partial #2 \partial #3}}
\def\bra#1{\left\langle #1\right|}
\def\ket#1{\left| #1\right\rangle}
\def\VEV#1{\left\langle #1\right\rangle}
\def\gdot#1{\rlap{$#1$}/}
\def\abs#1{\left| #1\right|}
\def\pr#1{#1^\prime}
\def\ltap{\raisebox{-.4ex}{\rlap{$\sim$}} \raisebox{.4ex}{$<$}}
\def\gtap{\raisebox{-.4ex}{\rlap{$\sim$}} \raisebox{.4ex}{$>$}}
\nwc{\Id}  {{\bf 1}}
\nwc{\diag} {{\rm diag}}
\nwc{\inv}  {{\rm inv}}
\nwc{\mod}  {{\rm mod}}
\nwc{\hal} {\frac{1}{2}}
\nwc{\tpi}  {2\pi i}
\def\contract{\makebox[1.2em][c]{
        \mbox{\rule{.6em}{.01truein}\rule{.01truein}{.6em}}}}
\def\slash#1{#1\!\!\!/\!\,\,}
%
%
\def\KK{{\rm I\kern -.2em  K}}
\def\NN{{\rm I\kern -.16em N}}
\def\RR{{\rm I\kern -.2em  R}}
\def\ZZ{Z \kern -.43em Z}
\def\QQ{{\rm \kern .25em
             \vrule height1.4ex depth-.12ex width.06em\kern-.31em Q}}
\def\CC{{\rm \kern .25em
             \vrule height1.4ex depth-.12ex width.06em\kern-.31em C}}
\def\ZZZ{Z\kern -0.31em Z}

\def \Msol {M_\odot}
\def\eV {\,{\rm  eV}}     
\def\KeV {\,{\rm  KeV}}     
\def\MeV {\,{\rm  MeV}}
\def\GeV {\,{\rm  GeV}}     
\def\TeV {\,{\rm  TeV}}     
\def\fm {\,{\rm  fm}}

\def\ap#1{Annals of Physics {\bf #1}}
\def\cmp#1{Comm. Math. Phys. {\bf #1}}
\def\hpa#1{Helv. Phys. Acta {\bf #1}}
\def\ijmpa#1{Int. J. Mod. Phys. {\bf A#1}}
\def\jpc#1{J. Phys. {\bf C#1}}
\def\mpla#1{Mod. Phys. Lett. {\bf A#1}}
\def\npa#1{Nucl. Phys. {\bf A#1}}
\def\npb#1{Nucl. Phys. {\bf B#1}}
\def\nc#1{Nuovo Cim. {\bf #1}}
\def\pha#1{Physica {\bf A#1}}
\def\pla#1{Phys. Lett. {\bf #1A}}
\def\plb#1{Phys. Lett. {\bf #1B}}
\def\pr#1{Phys. Rev. {\bf #1}}
\def\pra#1{Phys. Rev. {\bf A#1 }}
\def\prb#1{Phys. Rev. {\bf B#1 }}
\def\prp#1{Phys. Rep. {\bf #1}}
\def\prc#1{Phys. Rep. {\bf C#1}}
\def\prd#1{Phys. Rev. {\bf D#1 }}
\def\ptp#1{Progr. Theor. Phys. {\bf #1}}
\def\rmp#1{Rev. Mod. Phys. {\bf #1}}
\def\rnc#1{Riv. Nuo. Cim. {\bf #1}}
\def\zpc#1{Z. Phys. {\bf C#1}}
\def\APP#1{Acta Phys.~Pol.~{\bf #1}}
\def\AP#1{Annals of Physics~{\bf #1}}
\def\CMP#1{Comm. Math. Phys.~{\bf #1}}
\def\CNPP#1{Comm. Nucl. Part. Phys.~{\bf #1}}
\def\HPA#1{Helv. Phys. Acta~{\bf #1}}
\def\IJMP#1{Int. J. Mod. Phys.~{\bf #1}}
\def\JP#1{J. Phys.~{\bf #1}}
\def\MPL#1{Mod. Phys. Lett.~{\bf #1}}
\def\NP#1{Nucl. Phys.~{\bf #1}}
\def\NPPS#1{Nucl. Phys. Proc. Suppl.~{\bf #1}}
\def\NC#1{Nuovo Cim.~{\bf #1}}
\def\PH#1{Physica {\bf #1}}
\def\PL#1{Phys. Lett.~{\bf #1}}
\def\PR#1{Phys. Rev.~{\bf #1}}
\def\PRP#1{Phys. Rep.~{\bf #1}}
\def\PRL#1{Phys. Rev. Lett.~{\bf #1}}
\def\PNAS#1{Proc. Nat. Acad. Sc.~{\bf #1}}
\def\PTP#1{Progr. Theor. Phys.~{\bf #1}}
\def\RMP#1{Rev. Mod. Phys.~{\bf #1}}
\def\RNC#1{Riv. Nuo. Cim.~{\bf #1}}
\def\ZP#1{Z. Phys.~{\bf #1}}

\pagestyle{empty}
\noindent
\begin{flushright}
March 2003
\\
\end{flushright} 
\vspace{3cm}
\begin{center}
{ \Large \bf
The Quark-Meson Model\\ 
and\\
the Phase Diagram of Two-Flavour QCD \\
} 
\vspace{1cm}
{\Large 
N. Tetradis 
} 
\\
{\it
Department of Physics, University of Athens,
Zographou 157 71, Greece
} 
\\
\vspace{3cm}
\abstract{
We study the qualitative features of the
QCD phase diagram in the context of the
linear quark-meson model with two flavours, 
using the exact renormalization group.
We identify the universality classes of the second-order phase
transitions and calculate critical exponents. 
In the absence of explicit chiral-symmetry breaking
through the current quark masses, we 
discuss in detail the tricritical point 
and demonstrate that it is linked to the Gaussian fixed point.
In its vicinity
we study the universal crossover between the Gaussian and $O(4)$ fixed points,
and the weak first-order phase transitions.
In the presence of explicit chiral-symmetry breaking, we study in
detail the critical endpoint of the line of first-order phase transitions.
We demonstrate the decoupling of pion fluctuations and identify the
Ising universality class as the relevant one for the second-order phase
transition.
} 
\end{center}

\newpage

\pagestyle{plain}
\setcounter{page}{1}

\setcounter{equation}{0}

\section{Introduction}

The phase diagram of QCD at non-zero temperature and baryon
density has been the subject of many studies during the last years.
(For a review see ref. \cite{Rajagopal:2000wf}.) 
Parts of this phase diagram may be observable through the heavy-ion
collision experiments at RHIC and LHC. 
In particular, the conditions for the
transition from the hadronic phase to the quark gluon
plasma at high temperature and relatively low baryon density 
could be realized at these experiments. It is, therefore, desirable
to connect the various regions of the phase diagram with clear
experimental signatures. The most prominent feature of the phase diagram
in this respect is the critical endpoint that marks the end of the
line of first-order phase transitions. 
Its exact location and the size of the critical region around it 
determine its relevance for the experiments.
There exist several analytical studies of the phase diagram near the critical
endpoint. They employ effective theories of QCD
\cite{Asakawa:bq}--\cite{Kiriyama:2000yp}, as analytical calculations
in the context of the exact theory are very difficult. Lattice 
simulations are a source of precise information for full QCD. 
Recent studies 
\cite{Fodor:2001pe,deForcrand:2002ci,Allton:2002zi} have overcome
the difficulties associated with simulating systems with
non-zero chemical potential, even though the continuum limit or realistic 
quark masses 
have not been reached yet.

The investigation of large regions of the phase diagram, so as to capture 
the interplay
between critical and non-critical behaviour around the endpoint, seems 
more accessible to analytical methods. The description of
the critical region makes the use of the renormalization group 
necessary. The study of first-order phase transitions
is easier within the Wilsonian (or exact) formulation \cite{Wilson:1973jj}. 
For these reasons, we would like to derive the QCD phase
diagram employing an 
analytical approach that incorporates the exact renormalization group.
A lot of work has been done on this method
in recent years \cite{rome}.
We shall use the effective average action 
\cite{Wetterich:1989xg}, that results by integrating
out (quantum or thermal) fluctuations above a certain cutoff $k$.
Its dependence on 
$k$ is given by an exact evolution equation
\cite{Wetterich:yh,Tetradis:1993ts}, while
it becomes the effective action for $k=0$. 
It can by used in order to desribe very efficiently the universal and
non-universal aspects of second-order phase transitions, as well as
strong or weak first-order phase transitions \cite{Berges:2000ew}.

We shall not study the full QCD for several reasons: the 
various formulations of the exact renormalization group for gauge theories
\cite{Reuter:1993kw}--\cite{Morris:1999px}
have not been complemented yet with efficient approximations 
for practical calculations. Moreover, a
treatment of the exact evolution equation that could capture
the effects of the confining region of QCD has not been developed.

We shall work within an effective theory of QCD, the 
linear quark-meson model, that uses as fundamental
degrees of freedom mesons coupled to quarks. 
If only the two lightest quarks are taken into account the model is
similar to the $\sigma$-model of Gell-Mann and Levy \cite{Gell-Mann:np}, with
the nucleons replaced by quarks. The Langrangian involves four scalar
fields (the $\sigma$-field and the three pions $\pi_i$) arranged in 
a $2\times 2$ matrix $\Phi$, as well as the $u$ and 
$d$ quarks in a doublet. 
It is invariant under a global $SU(2)_L \times SU(2)_R$
symmetry acting on $\Phi$ and the quark doublet.
If the $\sigma$-field develops an expectation value the
symmetry is broken down to $SU(2)_{L+R}$. The model is very useful for
studying the spontaneous breaking of chiral symmetry, 
using $\Phi$ as the order 
parameter. The explicit breaking of this symmetry through the
quark masses can also be incorporated. It corresponds to the interaction
with an external source through a term $-j \sigma$ in the Lagrangian.

The linear quark-meson model 
has been studied for a general number of
flavours and its phenomenological viability has been
established \cite{Jungnickel:1995fp}.
Its formulation for non-zero temperature and quark chemical potential
has also been developed \cite{Berges:1998sd,Berges:1997eu}.
In particular, it has been found that, for two massless 
quark flavours, the model 
predicts a second-order phase transition for increasing temperature and
zero chemical potential, and a first-order one for increasing
chemical potential and zero temperature. 
It is likely, therefore, that the expected phase diagram of QCD for small
chemical potential 
can be reproduced correctly within this effective theory.
We point out, however, that the study of 
the colour superconducting or colour-flavour
locked phases \cite{Rajagopal:2000wf} requires 
the introduction
of degrees of freedom that correspond to diquark condensates.

In this work we shall study the phase diagram of
the quark-meson model.
We consider only the case of two flavours as the
basic structure appears already within
this approximation. The generalization to
three flavours is straightforward, even though the degree of
complexity increases significantly. We do not consider realistic
values for the various parameters of the Lagrangian. The reason is
that the treatment of the exact evolution equation 
becomes complicated for the realistic values and little analytical
progress can be made. Instead, one has to rely heavily on 
numerics that obscurs the basic features of the
problem. The consideration of parameter values 
consistent with the phenomenology is an obvious subject for
future studies.

Despite the above limitations, our work includes several 
important elements: The infrared region is
treated correctly through the renormalization group. 
The fixed point structure can be obtained in detail.
The universal behaviour near the critical points 
can be distinguished from the the non-universal aspects. 
The various universality classes can be identified and the crossover
behaviour from one to the other can be studied.

In section 2 we discuss the phase diagram in the context of mean field theory
in order to determine the main features we expect to appear.
In section 3 we present the main ingredients of the model we are considering.
In section 4 we derive the evolution equation that describes the
dependence of the potential on a scale $k$ that acts as an infrared cutoff.
In section 5 we cast this equation in a form that has no explicit dependence
on $k$ and facilitates the identification of fixed points.
In section 6 we derive the initial conditions for the solution of the
evolution equation at energy scales below the temperature, where 
the evolution becomes effectively three-dimensional.
In section 7 we use the formalism in order to discuss the phase diagram 
in the absence of 
explicit chiral symmetry breaking (zero current quark masses).
In section 8 we discuss the phase diagram in the presence of
explicit chiral symmetry breaking. 
In section 9 we summarize our findings and outline the directions of
future research.

\section{Mean field theory}

In order to obtain some intuition on the form of the phase diagram, it is
instructive to discuss a simple model in the mean-field approximation.
The Lagrangian contains standard kinetic terms for the fields 
$\sx$, $\pi_i$  ($i=1,2,3$), and the potential
\be
U(\rho)=m^2\rho+\frac{1}{2}\lx\rho^2+\frac{1}{3}\nu\rho^3,
\label{mpot} \ee
with $\rho=(\sx^2+\pi_i\pi^i)/2$. The couplings
$m^2$, $\lx$, $\nu$ are taken to be 
functions of two external parameters, which we
denote by $T$ and $\mu$ without having to be more specific at this stage.
We assume that $\nu$ is always positive, so that the potential is
bounded from below.

Let us consider first the case $\lx>0$. For
$m^2>0$ the minimum of the potential is located at $\sx=\pi_i=0$ and
the system is invariant under an $O(4)$ symmetry. 
For $m^2<0$ the minimum moves away from the origin. Without loss of
generality we take it along the $\sx$ axis. The symmetry is broken down
to $O(3)$. The $\pi$'s, which play the role of the
Goldstone fields, become massless at the minimum. 
If $m^2(T,\mu)$ has a zero at a certain value $T=T_{cr}$, while 
$\lx(T_{cr},\mu)>0$, the system undergoes a second-order 
phase transition at this point.
The minimum of the potential 
behaves as $\sx_0\sim |T-T_{cr}|^{1/2}$ slightly below the critical 
temperature. The critical exponent $\beta$ takes the mean field value
$\beta=1/2$.

Let us consider now the case $\lx<0$. 
It is easy to check that, as $m^2$ increases
from negative to positive values (through an increase of $T$ for example), 
the system undergoes a first-order
phase transition. For $|\lx|$ approaching zero the phase transition
becomes progressively weaker: The discontinuity in the order parameter 
(the value of $\sx$ at the minimum) approaches zero. For $\lx=0$ the
phase transition becomes second order. The minimum of the potential behaves
as $\sx_0\sim |T-T_{cr}|^{1/4}$. The critical exponent $\beta$ takes 
the value $\beta=1/4$ for this particular point.

Let us assume now for simplicity 
that $\lx$ is a decreasing function of $\mu$ only, 
and has a zero at $\mu=\mu_*$.
If we consider the phase transitions for increasing $T$ and fixed 
$\mu$ we find a line of second-order phase transitions for $\mu < \mu_*$,
and a line of first-order transitions for $\mu > \mu_*$.
The two lines meet at the
special point $(T_*,\mu_*)$, where $T_*=T_{cr}$ for $\mu=\mu_*$. This
point is characterized as a tricritical point.

If a source term $-j \sx$ is added to the potential of eq. (\ref{mpot})
the $O(4)$ symmetry is explicitly broken. The phase diagram is modified
significantly even for small $j$.
The second-order phase transitions, observed for 
$\lx >0$, disappear. The reason is that the minimum of the potential
is always at a value $\sx \not= 0$ that moves close to zero for increasing
$T$. For small $j$ the mass term at the minimum approaches zero at 
a value of $T$ near what we defined as $T_{cr}$ for $\lx=0$. However,
no genuine phase transition appears. Instead we observe an analytical
crossover. 

The line of first-order transitions 
persists for $j\not= 0$. However, 
for increasing $T$ the minimum jumps discontinuously between two non-zero
values of $\sx$ and never becomes zero. The line ends at a new special
point, whose nature can be examined by considering the $\sx$-derivative of
the potential $U(\rho)$ of eq. (\ref{mpot}). 
For a first-order phase transition to occur, 
$\partial U/\partial \sx$ must become equal to $j$ for three non-zero
values of $\sx$. This requires $\partial^2 U/\partial \sx^2 =0$
at two values of $\sx$. The endpoint corresponds to the situation that
all these values merge to one point. It can be checked that this requires
$\lx <0$. At the minimum $\sx_*$ 
of the potential $\Ut(\sx;j)=U(\sx,\pi_i=0)-j\sx$ 
at the critical endpoint, we have:
$d\Ut/d\sx=d^2\Ut/d\sx^2 =d^3\Ut/d\sx^3=0$. 
Therefore, at the endpoint we expect
a second-order phase transition for the deviation of $\sx$ from $\sx_*$.
As $\lx\not= 0$, the critical exponent $\beta$ takes the value $\beta=1/2$.
The $\pi$'s are massive, because
$m^2_\pi=\partial^2 U(\rho)/\partial \pi_i^2 = dU(\rho)/d\rho=j/\sx_*$ at the 
critical endpoint.

The purpose of the following is to recontruct the above picture 
in the context of the linear quark-meson model. We shall take into account
the effect of fluctuations of the fields, so that the potential will have
a more complicated form than the simple assumption of eq. (\ref{mpot}).
Moreover, the nature of the
fixed points will differ from the predictions of mean field theory.

\section{The model}

We consider an effective action of the form
\cite{Jungnickel:1995fp,Berges:1998sd,Berges:1997eu}

\begin{eqnarray}
  \Gm_k &=& \ds{
    \int^{1/T}_0 dx^0\int d^3x \Bigg\{  
    i \ovl{\psi}^a (\gamma^{\mu}\pa_{\mu} + \mu \gamma^0) \psi_a
    +h {\ovl{\psi}}^a \left[ \frac{1+\gamma^5}{2} {\Phi_a}^b
    - \frac{1-\gamma^5}{2} {(\Phi^{\dagger})_a}^b\right] \psi_b
    }\nonumber \\ 
  && \ds{
    \qquad\qquad\qquad \qquad \quad
    +\pa_{\mu}\Phi^*_{ab}\pa^{\mu}\Phi^{ab}
    +U_k(\rho;\mu,T)
    -\frac{1}{2}\left(\Phi_{ab}^*j^{ab}+j^*_{ab}\Phi^{ab} \right)
    \Bigg\} 
    }\, ,
  \label{truncation} 
\end{eqnarray}
with $\rho=\Tr\left(\Phi^\dagger \Phi\right)$. The field $\Phi$ includes
the $\sigma$ field and the three pion fields. It is parametrized as
\begin{equation}
  \Phi=\frac{1}{2}\left(\si+i\vec{\pi}\cdot\vec{\tau}\right)
  \label{phi}
\end{equation}
where $\tau_l$ $(l=1,2,3)$ denote the Pauli matrices. 
The fermionic field $\psi^a$ 
includes two flavours $(a=1,2)$ corresponding to the
up and down quarks. The non-zero temperature effects are taken into
account by restricting the 
time integration within the finite interval
$[0,1/T]$ in Euclidean space. The scalar (fermionic) fields obey 
periodic (anti-periodic) boundary conditions. We are also considering
a non-zero chemical potential, associated to the quark number density 
$\psi^\dagger_a \psi^a$ that is proportional to the baryon number.

In the effective action we have omitted the wavefunction renormalization 
for both the scalar and fermionic fields. We have also neglected the
scale dependence of the Yukawa coupling $h$. We take into account only
the scale dependence of the potential $U_k$, which we assume to be a
general function of $\rho$. 
Our approximation is consistent in two cases:\\
a) If the Yukawa coupling is small at an initial large momentum scale
$\Lx$, its subsequent evolution for $k<\Lx$ is expected to be slow.
The reason is that the leading contribution to $dh^2/d\ln k$ is $\sim h^2$
\cite{Berges:1997eu}. The same holds true for the wavefunction renormalizations
\cite{Berges:1997eu}. The scalar field wavefunction renormalization
receives an additional contribution through scalar field contributions.
Typically this is small in three and four dimensions, 
as is obvious from the smallness of
the anomalous dimension in the pure scalar theory.\\
2) If partial infrared fixed points exist in the evolution equations, all
the above quantities take almost constant values. The wavefunction 
renormalizations can be set equal to 1 through the rescaling 
of the fields, while the Yukawa coupling becomes scale independent. This
scenario is realized in the low energy regime of the quark-meson model in
four dimensions, for realistic values of the quark and meson masses. 

The last term in eq. (\ref{truncation}) accounts for the explicit breaking
of the chiral symmetry through the current quark masses. 
One expects $j \sim M = \diag(m_u,m_d)$.
We assume equal
masses $\hat m$ for the two light quarks so that
${j^a}_b \sim \hat{m} {\delta^a}_b$. 
The determination of
the proportionality constant requires the embedding of the 
theory described by the action (\ref{truncation}) in a more fundamental
framework. For example, in refs. 
\cite{Berges:1997eu,Ellwanger:wy,Jungnickel:1996fd} 
it was calculated
in terms of the energy scale $k_\Phi$ at which $\Phi$ can be
introduced as a composite field in a theory of quarks
(such as the Nambu--Jona-Lasinio model \cite{Nambu:tp}).
One obtains $j=2 Z_{\psi,k_\Phi} m^2_{k_\Phi} M/h_{k_\Phi}$, with 
$Z_{\psi,k_\Phi}$ the fermion wavefunction renormalization, and 
$m^2_{k_\Phi}$ the positive mass term in the potential of $\Phi$ at the
scale $k_\Phi$. This mass term becomes negative at lower energy scales
because of fermionic quantum fluctuations, and the chiral symmetry is 
spontaneously broken. In this work we set $Z_{\psi,k}=1$ for all $k$ and
also neglect the running of $h_{k}$. For this reason it is not possible to 
express $j$ in terms of fundamental quantities. However, $j$ can be 
fixed through the low energy dynamics. The pion
mass at the vacuum in
the presence of explicit chiral symmetry breaking is proportional
to the source term and can be used as input for its determination.

\section{The evolution equation}

The dependence 
of the effective action
$\Gamma_k$ on an infrared cutoff scale $k$ 
is given by an exact flow equation~\cite{Wetterich:yh}, which
for fermionic fields $\psi$ (quarks) and bosonic fields $\Phi$ (mesons)
reads
\begin{equation}
  \label{frame}
  \frac{\pa}{\pa k}\Gm_k[\psi,\Phi] = \frac{1}{2}{\rm Tr}_B 
  \left\{ \frac{\pa R_{kB}}
    {\pa k} \left(\Gm^{(2)}_k[\psi,\Phi]+R_k\right)^{-1}  \right\} 
    -{\rm Tr}_F \left\{ \frac{\pa R_{kF}}
    {\pa k} \left(\Gm^{(2)}_k[\psi,\Phi]+R_k\right)^{-1} 
  \right\} \, .
\end{equation}
Here $\Gamma_k^{(2)}$ is the matrix of second functional derivatives of
$\Gamma_k$ with respect to both fermionic and bosonic field components. The
first trace in the r.h.s. of~(\ref{frame}) runs only
over the bosonic degrees of freedom. It implies a momentum integration and
a summation over flavor indices. The second trace runs over the
fermionic degrees of freedom and contains in addition a summation over
Dirac and color indices. The terms $R_{kB}$, $R_{kF}$ are infrared cutoffs
that have been introduced in order to eliminate contributions from 
fluctuations with momenta $q^2\lta k^2$ in the momentum integrations.

The evolution equation for the scale dependent 
potential $U_k$ has been computed in ref. \cite{Berges:1998sd} starting
from eq. (\ref{frame}).
In the approximation that we neglect the running of the Yukawa 
coupling it takes the simplified form:
\begin{equation}
  \frac{\pa }{\pa k}U_k(\rho;T,\mu)=\frac{\pa }{\pa k}U_{kB}(\rho;T,\mu)
  +\frac{\pa }{\pa k}U_{kF}(\rho;T,\mu) \label{dtu} \, ,
\end{equation}
where
\begin{eqnarray}
  \ds{
    \frac{\pa }{\pa k}U_{kB}(\rho;T,\mu)} &=& \ds{
    \frac{1}{2} T \sum\limits_n
    \int\limits_{-\infty}^{\infty}
    \frac{d^3\vec{q}}{(2 \pi)^3} 
    \frac{\pa R_{kB}(\vec{q}^2+4n^2\pi^2 T^2)}{\pa k} \left[
    \frac{3}{P_{k}\left(\vec{q}^2+4n^2\pi^2 T^2\right) 
      + U_k'(\rho;T,\mu)}\right. 
     }\nonumber \\ 
  &&
  \ds{\qquad \qquad 
    +\left. \frac{1}{P_{k}\left(\vec{q}^2+4n^2\pi^2 T^2\right) 
   + U_k'(\rho;T,\mu)
      + 2 \rho U_k''(\rho;T,\mu)}\right]  
    \label{dtub} 
     }
\end{eqnarray}
and 
\begin{eqnarray}
  \ds{
    \frac{\pa }{\pa k}U_{kF}(\rho;T,\mu)} &=& 
     \ds{
     \frac{\pa }{\pa k} \left[ 
     {^4V}_{kF}(\rho)
     +{^3V}_{kF}(\rho;T,\mu)+{^3V}_{kF}(\rho;T,-\mu)
     \right]
     }\label{f1} \\ 
   \ds{{^4V}_{kF}(\rho)} &=&
    \ds{
    -8 N_c \frac{1}{2} 
    \int\limits_{-\infty}^{\infty}
    \frac{d^4q}{(2\pi)^4} \,\,\,\theta\left( \Lx^2-q^2\right) 
    \ln \left(q^{2}+k^2+h^2 \rho/2 \right) 
     }\label{f2} \\
   \ds{{^3V}_{kF}(\rho;T,\mu)} &=&
    \ds{
    -4 N_c T 
    \int\limits_{-\infty}^{\infty}
    \frac{d^3\vec{q}}{(2\pi)^3} 
    \ln \left(1 + \exp 
    \left[- \frac{1}{T}
     \left(\sqrt{\vec{q}^{\,2}+k^2+h^2 \rho/2}-\mu\right)\right]
    \right) 
     }\label{f3}  
     \, .
\end{eqnarray}
Here $\rho=\Tr\left(\Phi^\dagger \Phi\right)
=\left(\sx^2+\vec{\pi}^2 \right)/2$ 
and $N_c=3$ denotes the number of colours.  

The two terms in the r.h.s. of eq. (\ref{dtu}) correspond to the contributions
from scalar and fermionic fluctuations. The scalar term, given
by eq. (\ref{dtub}), includes the contributions from the radial mode 
(the $\sx$ field) and the three Godstone modes (the pions). 
The parametrization in terms of $\rho$ makes apparent the $O(4)$ symmetry
characterizing the mesonic sector in the two-flavour case.
The meson mass terms 
are $U_k'+2\rho U_k''$, $U_k'$, where primes denote
derivatives of the potential with respect to $\rho$.  
If the chiral symmetry is broken by an
expectation value for the $\sx$ field, the mass terms become
$m^2_\sx=\partial^2U_k/\partial \sx^2$, 
$m^2_\pi=(1/\sx)\partial U_k/\partial \sx$, as expected.

The scale dependent
propagators also contain the momentum dependent part
$P_{k}=q^2+R_{kB}(q^2)$. The term 
\be
R_{kB}(q^2)=\frac{q^2 e^{-q^2/k^2}}{1-e^{-q^2/k^2}}
\label{cutof} \ee
acts as an effective infrared cutoff in the loop integrations that
determine the effective potential. The scale dependence of $U_{kB}$ 
originates in this cutoff, as is apparent from the $k$-derivative of
$R_{kB}$ in eq. (\ref{dtub}). 
The periodic conditions imposed on the scalar fields for non-zero temperature
result in the replacement of the integration over $q^0$ by a sum over the
discrete Matsubara frequences. The chemical potential $\mu$ affects 
$U_{kB}$ only through the contribution of the fermionic part
$U_{kF}$ to the mass terms.

The fermionic contribution $U_{kF}$ has a simple interpretation within
our approximate treatment. 
It consists of three parts: The term ${^4V}_{kF}(\rho)$ originates in the 
quantum fluctuations of the fermionic fields, for which a mass-like infrared
cutoff $k^2$ has been introduced. This type of cutoff does
not provide automatic ultraviolet regularization. This is achieved 
through the modification of the cutoff by an additional $\theta$-function
\cite{Berges:1998sd}. The number of degrees of freedom is 
2(flavour)$\times$2(spin)$\times$2(particle-antiparticle)$\times N_c$
(colour).
The term ${^3V}_{kF}(\rho;T,\mu)$ is the free energy of a non-interacting
fermionic gas of temperature $T$ and chemical potential $\mu$
in the presence of the mass-like cutoff $k^2$. The ultraviolet 
regularization has been omitted as we assume $\Lx \gg T$.
The number of degrees of freedom is 2(flavour)$\times$2(spin)$\times N_c$
(colour).
The term ${^3V}_{kF}(\rho;T,-\mu)$ is the free energy of the anti-particles 
that have opposite chemical potential.
We emphasize that the evolution of $U_{kF}$ is derived starting from the
exact flow equation for the effective action of the system. 
The truncation of eq. (\ref{truncation}) and the additional assumptions
of constant wavefunction renormalizations and Yukawa coupling lead to
the intuitive form of eqs. (\ref{f1})--(\ref{f3}).
We also point out that the use of different cutoffs for the
scalar and fermionic fields does not affect physical
quantities, as the cutoffs are removed in the limit $k\to 0$.

\section{Scaling form of the evolution equation}

Rather than solving eq. (\ref{dtu}),
it is more convenient to solve the evolution equation for $U'_k$. This
can be written in the form
\begin{eqnarray}
    \frac{\pa }{\pa k}U'_{k}(\rho;T,\mu) &=&
    v_4 k \left(3U''_k+2\rho U'''_k\right)
      L^4_1\left( (U'_k+2\rho U''_k)/k^2;T\right)
     +3v_4\, k \,U''_k \,L^4_1\left(U'_k/k^2;T\right)
\nonumber \\
&&    +\frac{\pa^2 }{\pa k \pa \rho}      \left[ 
     {^4V}_{kF}(\rho)
     +{^3V}_{kF}(\rho;T,\mu)+{^3V}_{kF}(\rho;T,-\mu)
     \right] \, ,
    \label{eveq} \end{eqnarray}
where, in general dimensions $d$,
\be
v_d^{-1}=2^{d+1} \pi^{d/2} \Gamma \left( \frac{d}{2} \right).
\label{twofive} \ee 
The ``threshold'' functions $L^4_1$
are particular cases of 
\be
L^d_n(w;T) = 
- 2 n k^{2n-d+1} \pi^{-\frac{d}{2}+1} \Gamma \left( \frac{d}{2} \right) 
 T \sum_m
\int d^{d-1} \vec{q}~~ \frac{\partial P_{k}}{ \partial k} 
(P_{k} + w\,k^2)^{-(n+1)}, 
\label{fourfive} \ee
where $P_{k}$ is a function of $\vec{q}^2+4n^2\pi^2 T^2$, as we 
discussed earlier.

The functions $L^d_n(w;T)$ have been discussed extensively in refs. 
\cite{Tetradis:1992qt,Tetradis:1992xd,Tetradis:1993ts,Berges:2000ew}.
Their basic properties can be established
analytically. 
For $T \ll k$ the summation over discrete 
values of $m$ in the expression (\ref{fourfive}) is equal to the 
integration over a continuous range of $q^0$, up to exponentially small 
corrections.
Therefore 
\be
L^d_n(w;T) = L^d_n(w,0)\equiv L^d_n(w)~~~~~~~~~{\rm for}~~T \ll k.
\label{foursix} \ee
In the opposite limit $T \gg k$ the summation over $m$ is
dominated by the $m=0$ contribution. 
This results in the simple expression
\be 
L^d_n(w;T) = \frac{v_{d-1}}{v_d} \frac{T}{k} L^{d-1}_n(w)~~~~~~~~~~~~{\rm 
for}~~T \gg k,
\label{fourseven} \ee
with $v_d$ defined in eq. (\ref{twofive}).
The two regions of $T/k$ in which $L^d_n(w;T)$ is given by the 
equations (\ref{foursix}), (\ref{fourseven}) 
are connected by a small interval $T/\theta_2<k<T/\theta_1$ with a more 
complicated dependence on
$w$ and $T$. The transitions at $k=T/\theta_1$ and $k=T/\theta_2$ are
sharp for an exponential cutoff, such as the one given by eq. (\ref{cutof}).
The functions $L^d_n(w)$ fall off 
for large values of $w$, following a power law. As 
a result they introduce a threshold behaviour for the 
contributions of massive modes to the evolution equations. 
These
include $L^d_n$ functions with the mass eigenvalues divided by $k^2$
as their arguments.
When the running squared mass of a massive mode 
becomes much larger than the scale $k^2$, 
the mode decouples the respective contribution vanishes.
We evaluate the expressions for $L^d_n(w;T)$ 
numerically and use numerical fits for the solution of 
the evolution equations.

The fermionic contribution is expected to display a similar behaviour
for decreasing $k$. It can be checked that 
the temperature-dependent term of eq. (\ref{f3})
evolves slowly for $k\lta T$ and obtains asymptotically a 
constant form. However, the $k$ dependence disappears only as power law,
instead of exponentially fast as for $L^d_n(w,T)$ above.
The quantum contribution of eq. (\ref{f2}) 
has a strong dependence on the ultraviolet cutoff $\Lx$ that regulates
the momentum integration. As a result, the quantity 
$\partial^2\, (^4V_{kF})/\partial k \partial \rho$ displays a logarithmic
dependence on $\Lx$.
The above shortcomings are connected to the use of 
a mass-like cutoff for the fermionic sector, instead of an exponential
cutoff as for the bosons. A mass like cutoff is the simplest choice that
preserves the Lorentz structure of the kinetic term of free fermions
for non-zero chemical potential, so as to make
practical calculations feasible \cite{Berges:1998sd}.
A more ingenuous, and more complicated, form is required in order to
preserve the correct Lorentz structure and guarantee exponential
decoupling of the temperature effects for fermions at low energy scales.

In this work we are not interested in the
details of fermion decoupling. For this reason we 
follow a different approach. We assume that at a sufficiently
low scale $k\lta T$ the fermionic contributions to the evolution
equation decouple. This is the expected behaviour, as the lowest Matsubara
frequency is $\pi T$ for the fermions and no zero mode exists.
The result of the integration of the fermionic modes is a contribution
equal to the perturbative one-loop term for the potential. 
It is clear that this assumption is consistent with the form of 
the terms of eqs. (\ref{f2}), (\ref{f3}) for $k=0$. It is expected to 
be correct for small Yukawa coupling $h$.

We concentrate on the behaviour of the evolution equation 
(\ref{eveq}) for $k\lta T$, in which, as we explained above,
two important simplifications
occur: \\
a) The ``threshold'' functions are approximated very well by
$L^4_1(w,T)=(4T/k) L^3_1(w)$. \\
b) The fermions have decoupled because of their thermal masses.\\
We define the quantities
\begin{eqnarray}
{^3U}_k(\rho_3)&=& \frac{U_k(\rho;T,\mu)}{T}
\label{ut} \\
\rho_3&=&\frac{\rho}{T},
\label{rhot} \end{eqnarray}
and their dimensionless versions
\begin{eqnarray}
u_k(\rht)&=&\frac{{^3U}_k(\rho_3)}{k^3}=\frac{{U}_k(\rho;T,\mu)}{k^3T}
\label{utt} \\
\rht&=&\frac{\rho_3}{k}=\frac{\rho}{kT}.
\label{rhott} \end{eqnarray}
The evolution equation (\ref{eveq}) can be rewritten in the scaling form
\cite{Tetradis:1993ts}
\be
    \frac{\pa }{\pa t}u'_{k}(\rht) =
      -2 u'_k+\rht u''_k  +v_3 \left(3u''_k+2\rht u'''_k\right)
      L^3_1\left( u'_k+2\rht u''_k\right)
     +3v_3\, u''_k \,L^3_1\left(u'_k\right),
\label{scinv} \ee
with $t=\ln(k/M)$ and the primes denoting derivatives with respect to
$\rht$. The energy scale $M$ can be defined arbitrarily.
The characteristic property of the above equation is that the scale 
$k$ does not appear explicitly in it. This permits the search for
fixed points, i.e. scale independent solutions. They are 
the solutions with $\partial u'_k/\partial t=0.$

The evolution equation (\ref{scinv})
has a form typical of a three-dimensional theory (dimensional reduction).
This is expected, as only the fields with zero Matsubara frequences 
(the scalar zero modes) are sufficiently light in order to contribute
to the evolution.

\section{Initial conditions for the evolution equation}

The initial conditions for the solution of eq. (\ref{scinv})
must be given at the scale $k_T=T/\theta_2$.
Let us assume that at the large momentum scale $\Lx$ (where the effects
associated with the temperature and chemical potential are negligible)
the potential has the simple form 
\be
U_\Lx(\rho)=U_\Lx(\rho;0,0)=
\frac{\lx}{2} \left(\rho -\rho_{0\Lx}\right),
\label{cut} \ee
with $\rho_{0\Lx}<0$.
For small $\lx$ we can integrate the evolution equation (\ref{eveq})
between the scales $\Lx$ and $k_T=T/\theta_2$, approximating the arguments
of the threshold functions with zero and setting $U''_k=\lx$, $U'''_k=0$.
These approximations account for the leading pertubative 
result and give \cite{Tetradis:1992xd,Tetradis:1996fw}
\begin{eqnarray}
U'_{k_T}(\rho;T,\mu) & \simeq & \lx \left[
\rho-\left( \rho_{0\Lx} - \frac{3}{16\pi^2} \Lx^2\right)-\frac{3}{4\pi^2} T^2 
\left( 
\frac{\sqrt{\pi}}{\theta_2} -\frac{\pi^2}{3} \right)
\right] 
\nonumber \\ 
&& +{^4V'}_{0F}(\rho)+{^3V'}_{0F}(\rho;T,\mu)+{^3V'}_{0F}(\rho;T,-\mu).
\label{init2} 
\end{eqnarray}
with ${^4V}_{0F}$, ${^3V}_{0F}$ given by eqs. (\ref{f2}), (\ref{f3}) with
$k=0$.
We have assumed that the fermionic contributions have already 
decoupled at $k_T=T/\theta_2$
and are given by eq. (\ref{f3}) with $k=0$.
The term $\sim T^2$ results from 
the complicated form of $L^4_1(0;T)$ in the $k$-interval
$[T/\theta_2,T/\theta_1]$ \cite{Tetradis:1996fw}.

The $T=\mu=0$  fermionic contribution (\ref{f2}) 
can be evaluated explicitly
\be
{^4V'}(\rho)=-N_c\frac{h^2}{8\pi^2}\Lx^2
-N_c\frac{h^4}{16\pi^2}\rho \ln \left( 
\frac{\frac{h^2\rho}{2\Lx^2}}{1+\frac{h^2\rho}{2\Lx^2}}\right).
\label{vf0} \ee 
We can cast eq. (\ref{init2}) in the form 
\be
U'_{k_T}(\rho;T,\mu) \simeq  \lx \left[
\rho-\rho_{0R}-\frac{3}{4\pi^2} T^2 
\left( 
\frac{\sqrt{\pi}}{\theta_2} -\frac{\pi^2}{3} \right)
\right] +I_F(\rho;T,\mu)
\label{init} \ee
with
\begin{eqnarray}
\rho_{0R}&=&\rho_{0\Lx} - \frac{3}{16\pi^2} \Lx^2
+\frac{h^2}{\lx}\frac{N_c}{8\pi^2}\Lx^2
\label{rhoren} \\
I_F(\rho;T,\mu)&=&
-N_c\frac{h^4}{16\pi^2}\rho \ln \left( 
\frac{\frac{h^2\rho}{2\Lx^2}}{1+\frac{h^2\rho}{2\Lx^2}}\right)
+{^3V'}_{0F}(\rho;T,\mu)+{^3V'}_{0F}(\rho;T,-\mu).
\label{if} \end{eqnarray}

The $T=\mu=0$ potential 
displays the spontaneous chiral symmetry breaking
induced by the fermionic fluctuations.
For sufficiently large Yukawa coupling $h$, 
even if $\rho_{0\Lx}$ is negative (corresponding to a positive mass term),
the renormalized value $\rho_{0R}$ can become positive through the 
fermionic contribution in eq. (\ref{rhoren}).

At non-zero temperature,  
a consistency check for the initial condition (\ref{init})
can be performed by integrating eq. (\ref{scinv}) from $k=T/\theta_2$ to
$k=0$.
Again we assume
a small $\lx_R$, set $U''_k=\lx_R$, $U'''_R=0$ 
and take the arguments of the ``threshold'' functions equal
to zero. In this way we obtain
\be
U'_0(\rho;T,\mu) 
\simeq
U'_{k_T}(\rho;T,\mu) +
\frac{3}{4\pi^2}\frac{\sqrt{\pi}}{\theta_2}\lx T^2
=
\lx \left[
\rho-\rho_{0R}+6\frac{T^2}{24}
\right] +I_F(\rho;T,\mu),
\label{check} \ee
where we have used $L^3_1(0)=-\sqrt{\pi}$.
This expression reproduces correctly the leading perturbative result
\cite{Dolan:qd} for the critical temperature. 
However, the assumption that the couplings remain
small is not valid near the phase transiton. Close to
the critical temperature the fixed-point
structure of the theory becomes important and the arguments of the
``threshold'' functions cannot be neglected.

Another ingredient that we have neglected in our considerations so far is the
explicit chiral symmetry breaking term $\sim \sx$, arising through the
current quark masses. A term linear in $\sx$ does not appear expicitly
in the evolution equations, as the flow equation (\ref{frame})
involves second functional derivatives of the action with respect to the
fields. For this reason the linear term can be added directly to the
potential at any scale. 


\section{The phase diagram for $j=0$}

Let us summarize the main points of the framework we have set up.
Our theory is defined through the potential of eq. (\ref{cut})
at a scale $\Lx \gg T$. 
In a realistic theory we can identify $\Lx$ with the
energy scale $k_\Phi$ at which the mesonic fields $\sx$, $\pi_i$ can be
introduced as composite fields in a theory of quarks.
The renormalized potential for $k=0$ and zero temperature and chemical
potential is given by
eqs. (\ref{init})--(\ref{if}) with $T=\mu=0$. These approximate 
expressions are correct if: \\
a) $h$ is small enough for its logarithmic running to be neglected. This
requires corrections
$\sim h^4/(4\pi^2)$ to be negligible relative to $h^2$.\\ 
b) $\lx$ is small enough for terms $\sim \lx^2$ to be neglected.
Notice, however, that corrections $\sim h^4/(4\pi^2)$ are not assumed
to be small relative to $\lx$. On the contrary, it is for the range 
$h^4/(4\pi^2)\sim \lx$ that 
first-order phase transitions appear. \\
The explicit chiral symmetry breaking can be taken into account by
adding a term $-j\sx$ to the potential. 
The parameters $\rho_{0R}$, 
$\lx$, $j$ can be determined through the 
pion decay constant $f_\pi$ and the masses 
$m_\sx$, $m_\pi$ of the $\sx$ and pion fields.
The Yukawa coupling $h$ can be determined through the constituent quark
mass $M_q$.

At the scale $k_T=T/\theta_2$ the potential is 
given by eq. (\ref{init2}), with
${^3V}_{0F}$ given by eq. (\ref{f3}) for
$k=0$.
The evolution of $U_k$ for 
$k\leq k_T$ is given
by eq. (\ref{scinv}). The scalar modes with non-zero Matsubara frequencies,
as well as the fermions, have completely decoupled. Only the
fluctuations of the scalar zero modes contribute to the evolution of
the potential. For this reason the theory is effectively three-dimensional
at these low scales. 
For practical calculations we use $\theta_2=1$.

The fermionic contribution $I_F(\rho;T,\mu)$ to the initial condition 
for the potential, given by eq. (\ref{if}), can be evaluated in certain
limiting cases for small mass term $h^2\rho/2$. For $\mu=0$ 
the standard high temperature expansion \cite{Dolan:qd} gives
\be
I_F(\rho;T,0) = N_c \frac{h^2}{12} T^2 
+N_c \frac{h^4}{16\pi^2} \rho 
\ln \left( \frac{\Lx^2}{\alpha T^2}\right)
+{\cal O}( h^6),
\label{mu0} \ee
with $\alpha\simeq 8.5$.
For $T=0$ the Fermi-Dirac distribution becomes a step-function
and we obtain
\be
I_F(\rho;0,\mu) = N_c \frac{h^2}{4\pi^2} \mu^2 
+N_c \frac{h^4}{16\pi^2} \rho 
\ln \left( \frac{\Lx^2}{\beta \mu^2}\right)
+{\cal O}( h^6),
\label{T0} \ee
with $\beta\simeq 10.9$.

The terms $\sim h^4 \rho$ in the r.h.s. of the above equations can
change the sign of the tree-level
term $\lx \rho$ in the initial condition (\ref{init}).
The phase diagram expected for QCD is obtained if the total 
term $\sim \rho$ is positive near the phase transition
for $T\not=0$, $\mu=0$ and negative
for $T=0$, $\mu\not=0$. In our model we choose couplings such that 
this condition is satisfied. It has been checked explicitly 
\cite{Berges:1998sd,Berges:1997eu} that the 
realistic values of the parameters, consistent with low-energy
QCD phenomenology, reproduce the correct order for the transitions
for $T\not=0$, $\mu=0$ and 
$T=0$, $\mu\not=0$. 

We define dimensionful quantities in terms of the parameter $\rho_{0R}$
defined in eq. (\ref{rhoren}).
Our analysis of the phase diagram will be carried out for a model
with $\lx=0.1$, $h^2=2.3$, $\Lx^2=1.8\,\rho_{0R}$.

\subsection{The second-order phase transitions}

\begin{figure}[t]
 \centerline{\epsfig{figure=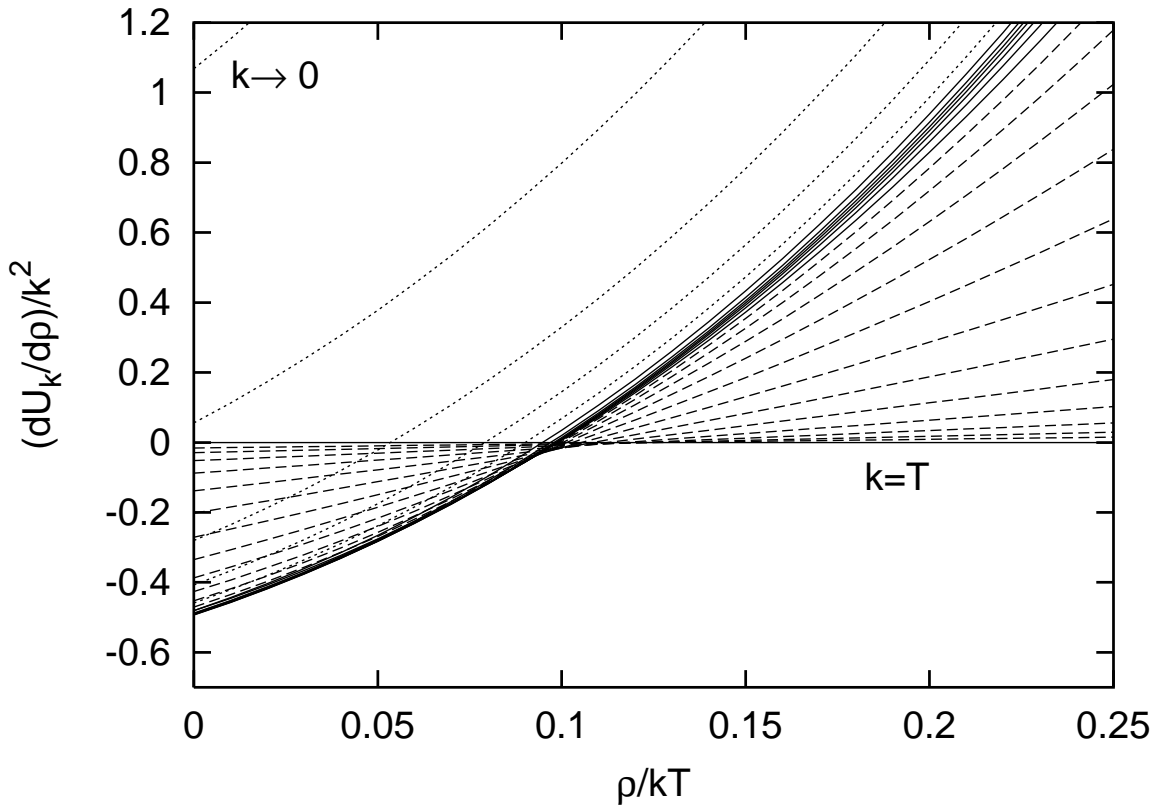,width=11cm,angle=-90}}
 \caption{\it
The evolution of the rescaled potential for 
$\mu=0$, $T/\sqrt{\rho_{0R}}\simeq 0.407.$
}
 \label{fig1}
 \end{figure}

We start by considering a system with zero chemical potential.
In fig. \ref{fig1} we display the solution of eq. (\ref{scinv}),
for a theory with 
$\mu=0$, $T/\sqrt{\rho_{0R}}\simeq 0.407.$
We start the evolution at a scale $k_T=T$ and move towards the infrared.
The initial condition is given by eqs. (\ref{init}), (\ref{if}), and
corresponds to a system in the broken phase.
We observe that initially $u_k'(\rht)$ is small, because of the
smallness of the value of $\lx$ we chose. Subsequently the 
potential evolves (dashed lines), while its minimum $\rht_0$ remains
non-zero. At a later stage the solution approaches an almost constant
form (solid lines). 
This is the fixed point of the $O(4)$ universality class.
At the last stages of the evolution (dotted lines) the minimum of the 
potential moves to the origin and $u_k'(0)$ becomes positive. 

For $k\to 0$,
the rescaled mass term $u_k'(0)$ diverges so that 
$dU_k(0;T)/d\rho=k^2du_k(0)/d\rht$ approaches a constant value. The system
ends up in the symmetric phase.  
The renormalized mass term $m^2_R(T)=dU_0(0;T)/d\rho$ 
is very small in terms of the 
characteristic mass scale of the potential at $k=T$. The reason is that,
during the ``time'' $t=\ln(k/\rho^{1/2}_{0R})$ 
the solution is close the fixed point, the quantity 
$u'_k(0)$ remains constant, so that $U'_k(0;T)$ scales $\sim k^2$.
The longer the system spends near the fixed point,
the smaller the final value of $m^2_R(T)$.
The approach to the fixed point is controlled by one
parameter, the temperature $T$. For the evolution displayed in fig.
\ref{fig1} the temperature is slightly larger than the critical value 
$T_{cr}$ that
separates the symmetric from the broken phase. 
For a value slightly smaller than the critical, the evolution would be 
similar to that in fig. \ref{fig1}, apart from the last stages. 
For $k\to 0$ the minimum $\rht_0$ of $u_k(\rht)$ would diverge, so that
the minimum $\rho_0$ of $U_k(\rho;T)$, given by $\rho_0(k,T)=\rht(k) k T$,
would reach a constant very small value $\rho_{0R}(T)$. 
In this way, the system would
end up in the broken phase.
Examples of this behaviour can be found in the review \cite{Berges:2000ew} 
and references therein.

\begin{figure}[t]
 \centerline{\epsfig{figure=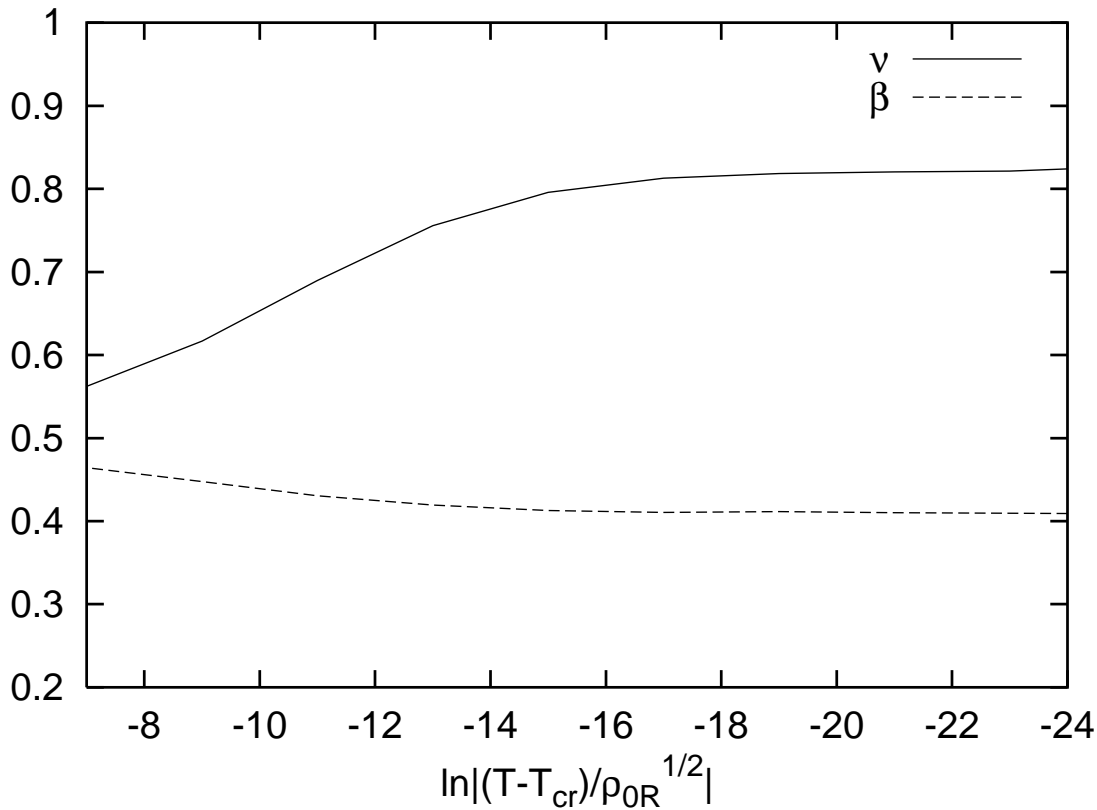,width=11cm,angle=-90}}
 \caption{\it
The critical exponents $\nu$ and $\beta$ as the critical temperature
is approached.}
 \label{fig2}
 \end{figure}

\begin{figure}[t]
 \centerline{\epsfig{figure=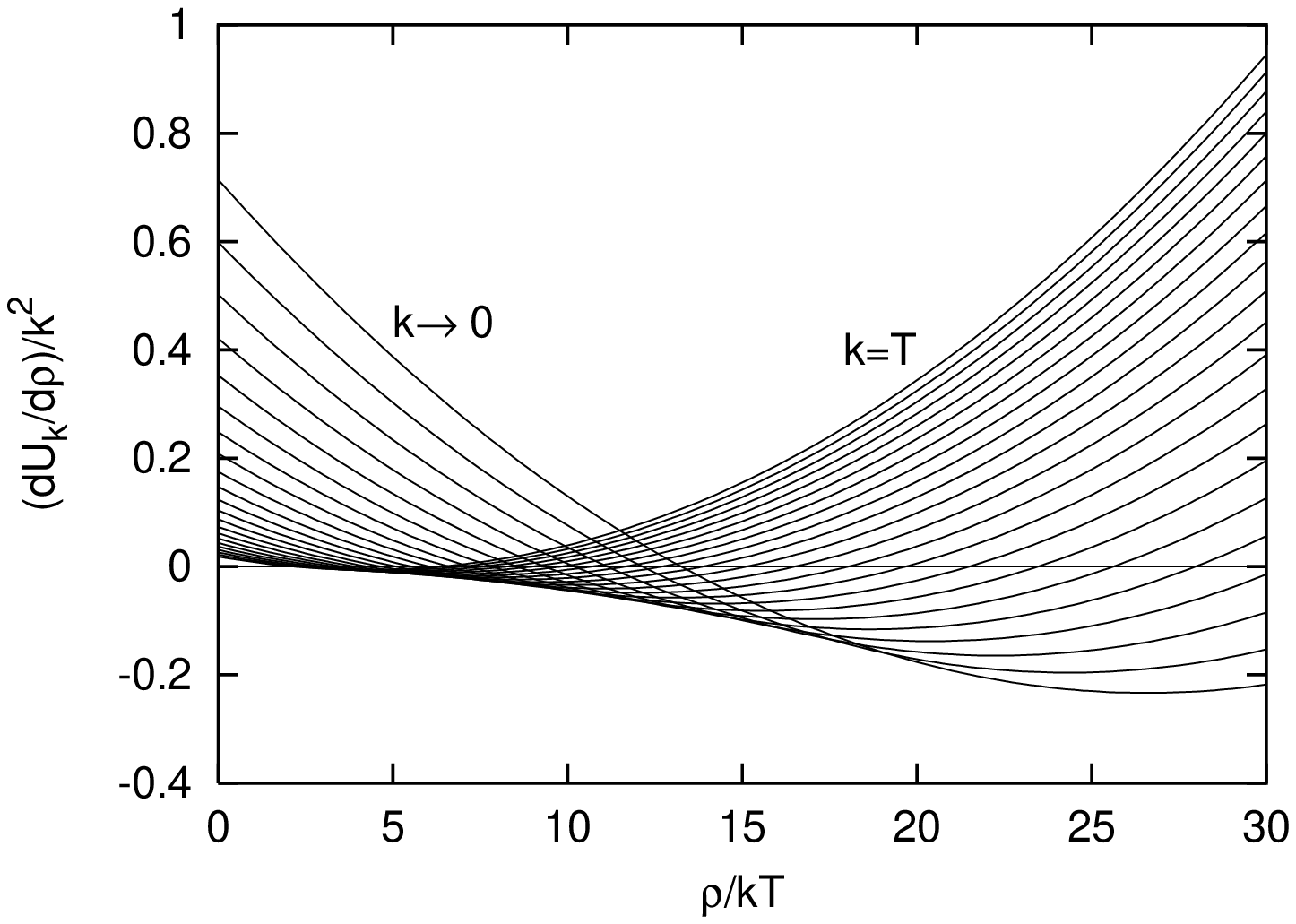,width=11cm,angle=-90}}
 \caption{\it
The evolution of the rescaled potential for 
$\mu/\sqrt{\rho_{0R}} \simeq 0.735$, $T/\sqrt{\rho_{0R}}=0.1.$
}
 \label{fig3}
 \end{figure}

The fixed point of the $O(4)$ universality class is approached by fine-tuning
only one relevant parameter, the temperature $T$ of the system.
The temperature effectively controls the mass term of the initial potential.
The properties of the critical theory (the one for $T\simeq T_{cr}$)
are determined only by the fixed point. The reason is that the system,
after spending a significant part of the evolution near the fixed point,
loses memory of the initial conditions. 
As a result, all physical quantities at the vacuum can be parametrized
in terms of the deviation of $T$ from $\tcr$. The dependence on
$T-\tcr$ is not analytic. In particular, for the mass term and the 
expectation value we expect a behaviour that can be parametrized as
\begin{eqnarray}
m^2_R(T)&\sim&\left|T-\tcr \right|^{2\nu}
\label{nnu} \\
\rho_{0R}(T)&\sim&\left|T-\tcr \right|^{2\beta}.
\label{bbeta} \end{eqnarray}
In fig. \ref{fig2} we plot the effective values of the critical 
exponents $\nu$, and $\beta$, extracted through the integration of the
evolution equation for various values of $T$. We observe that the exponents
become constant very close to the critical temperature, with values
$\nu\simeq 0.82$, $\beta\simeq 0.41$. These are in reasonable agreement with
the values determined through high precision calculations using alternative
methods, or the effective average action taking into account the 
wavefunction renormalization: 
$\nu\simeq 0.74$, $\beta\simeq 0.38$ \cite{Berges:2000ew}.
The descrepancy arises because we have neglected the wavefunction
renormalization, effectively using $\eta=0$ for the anomalous dimension
of the field. The correct value of this quantity is 
$\eta\simeq 0.036$. Notice, however, that the scaling law 
$\beta=(1+\eta)\nu/2$ is satisfied by our result.

\subsection{The first-order phase transitions}

We turn now to the region of first-order phase transitions, which are
expected to take place for large values of the chemical potential. 
We consider a system with 
$\mu\simeq 0.735$, $T/\sqrt{\rho_{0R}}= 0.1.$
In fig. \ref{fig3} we display the evolution of $u'_k(\rht)$
starting at $k=T$ and moving towards $k=0$.
This figure should be compared to fig. \ref{fig1}.
The differences are significant. Already at $k=T$, 
there are two values of the order
parameter $\rht$ for which $u'_k$ vanishes. Moreover, 
$u'_k(0)$ is positive. This implies a potential 
with two minima (one located at the origin) and a maximum.
During the evolution towards the infrared the system does not approach
a fixed-point solution. For $k\to 0$, $u'_k(0)$ increases so that
$U'_k(0;T,\mu)$ reaches a constant value. The point
$\rht_0$ where the second minimum of $u_k$
is located diverges also, so that the second minimum $\rho_0$
of $U_k$ approaches a constant value.

\begin{figure}[t]
 \centerline{\epsfig{figure=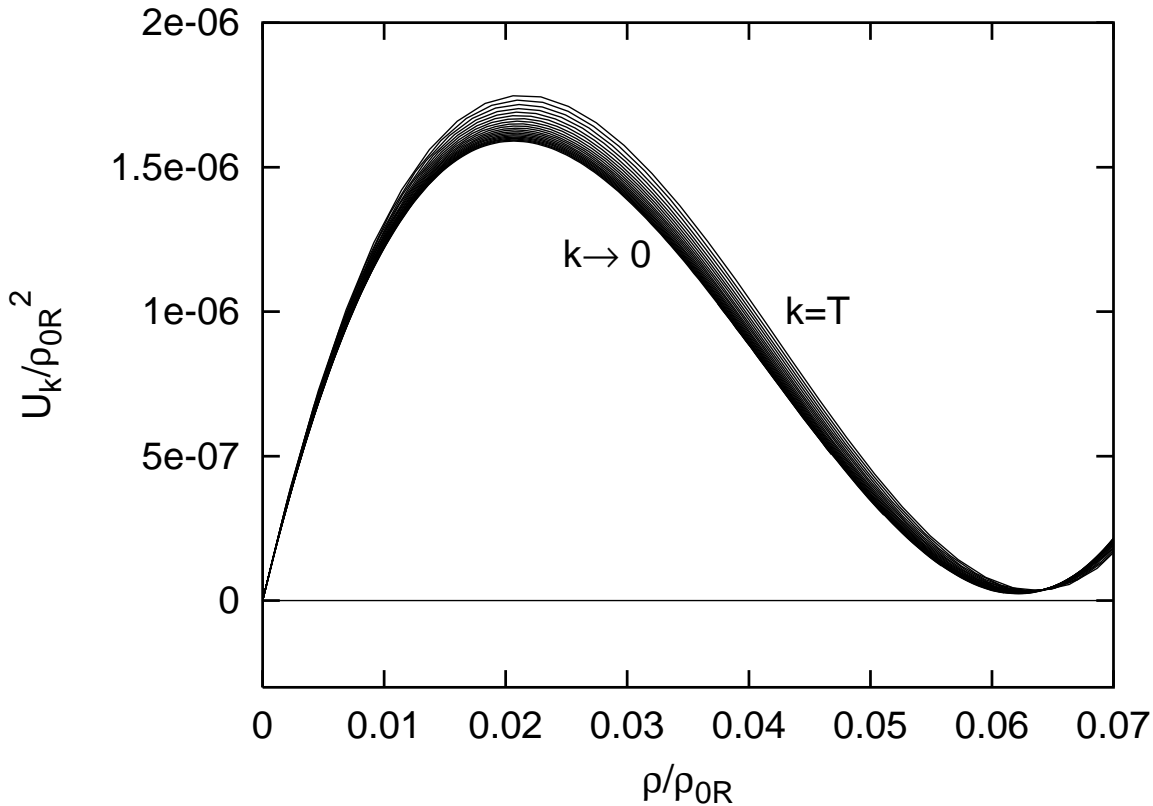,width=11cm,angle=-90}}
 \caption{\it
The evolution of the potential for 
$\mu/\sqrt{\rho_{0R}}\simeq 0.735$, $T/\sqrt{\rho_{0R}}= 0.1.$
}
 \label{fig4}
 \end{figure}

The behaviour of $U_k(\rho;T,\mu)$ is displayed in fig. \ref{fig4}.
We observe the two minima, and the curvature of the
potential around them that determines the mass terms,
settling down at constant values. The minimum at the origin is slightly 
deeper than the one at non-zero $\rho$, indicating that the temperature is
slightly larger than the critical one.  
The barrier separating the two minima remains scale dependent 
even after the minima have become scale independent to a good approximation.
This is a general feature of the solutions near first-order phase
transitions, which is related to the issue of the convexity of the
effective potential \cite{Tetradis:1992qt,Berges:2000ew}. 
As the scale-dependent
potential $U_k$ approaches the effective potential for $k\to 0$, the
barrier should disappear, giving its place to the standard Maxwell
construction. This is a general property of the solutions of the evolution
equation \cite{Tetradis:1992qt,Berges:2000ew}, which can be 
established analytically. However, the numerical integration of the 
evolution equation in this regime is difficult because of the
presence of a pole in the threshold function. For this reason, 
in fig. \ref{fig4} we have stopped the evolution at a non-zero value of $k$, 
such that the minima are stabilized, even though the barrier has not
disappeared yet. An example of 
a more elaborate numerical integration that displays the
approach to convexity for $k\to 0$ is presented in ref. \cite{Berges:2000ew}. 
The calculation of bubble-nucleation rates during a first-order phase
transition is performed at non-zero values of $k$. The physical value of
the nucleation rate is independent of the choice of $k$ at which the 
calculation is carried out \cite{Strumia:1998nf}.

\begin{figure}[t]
 \centerline{\epsfig{figure=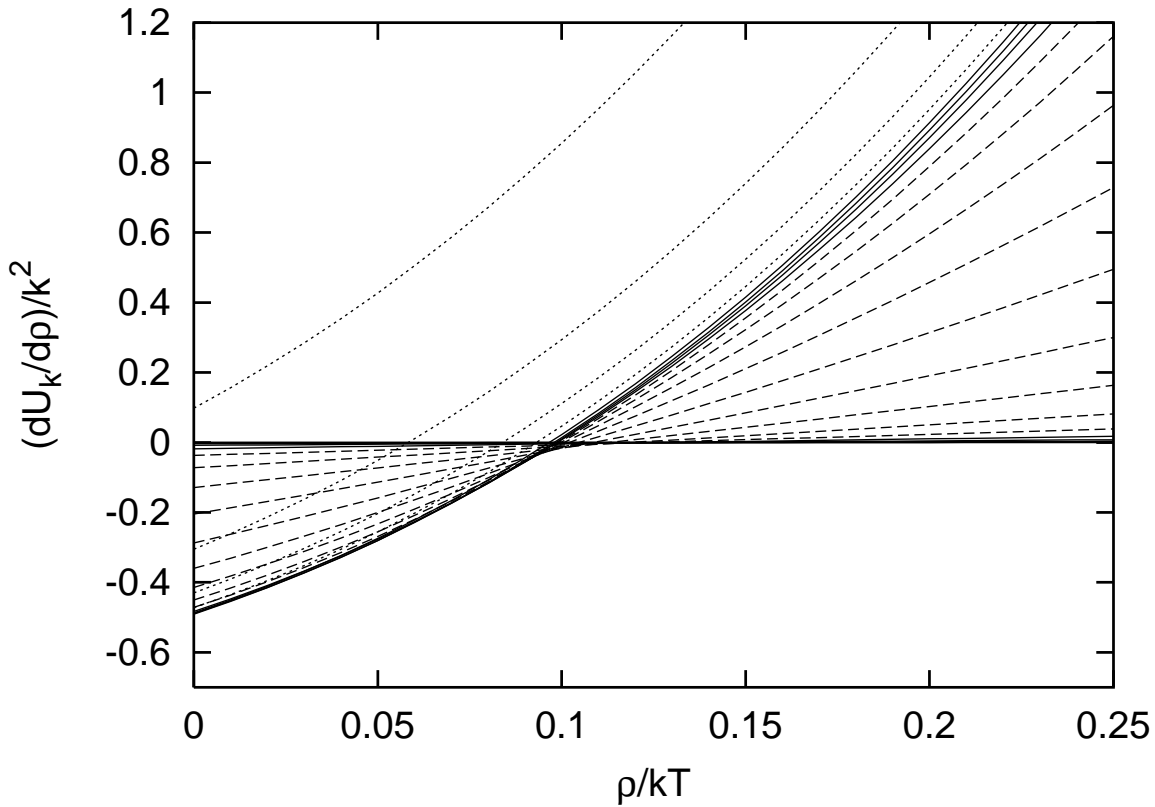,width=11cm,angle=-90}}
 \caption{\it
The evolution of the rescaled potential for 
$\mu/\sqrt{\rho_{0R}}=0.718$, $T/\sqrt{\rho_{0R}}\simeq 0.130.$
}
 \label{fig5}
 \end{figure}

\subsection{The tricritical point}

The second-order phase transitions, discussed in subsection 7.1, form a line
on the phase diagram, which starts at $\mu=0$ and continues towards
non-zero values of $\mu$. Similarly, the first-order phase transitions, 
discussed in subsection 7.2,
form another line that starts at $T=0$ and continutes towards non-zero values
of $T$. These two lines are expected to meet at a point on the
($T$,$\mu$) plane, characterized as the 
tricritical point. Near it the first-order transitions are very
weak (the discontinuity in $\rho$ approaches zero) and eventually become
second order. 

In fig. \ref{fig5} we display the evolution of $u'_k(\rht)$ for
$\mu/\sqrt{\rho_{0R}}=0.718$, $T/\sqrt{\rho_{0R}}\simeq 0.130.$
This figure is very similar to fig. \ref{fig1}. The potential evolves 
towards the $O(4)$ fixed point, where it becomes stationary for a while.
Eventually it moves away from the fixed point and settles down in the
symmetric phase. 

The significant difference between the two figures concerns the initial 
stage of the evolution. In fig. \ref{fig1} this is rather short, as the
potential moves quickly away from its initial form and approaches the 
fixed point. In fig. \ref{fig5} the initial stage is much longer, as can
be deduced from the accumulation of curves on the horizontal axis. In fact,
two fixed points are apparent in fig. \ref{fig5}: the one characterizing
the $O(4)$ universality class, and the Gaussian fixed point which corresponds
to $u'_k(\rht)=0$. 
The system starts very close to the Gaussian fixed point. This
is unstable, so that eventually the system approaches the $O(4)$ fixed point.
In order to start near the Gaussian fixed point, two relevant parameters must
be fine-tuned: the temperature and the chemical potential. They control the
mass term (the term $\sim \rho$) and the quartic coupling (the term
$\sim \rho^2$) 
of the initial potential. In particular,
the initial values of these two terms are very close to zero, so that the
term $\sim \rho^3$  (in an expansion of $U_{k_T}(\rho)$ around the
origin) is important. 

\begin{figure}[t]
 \centerline{\epsfig{figure=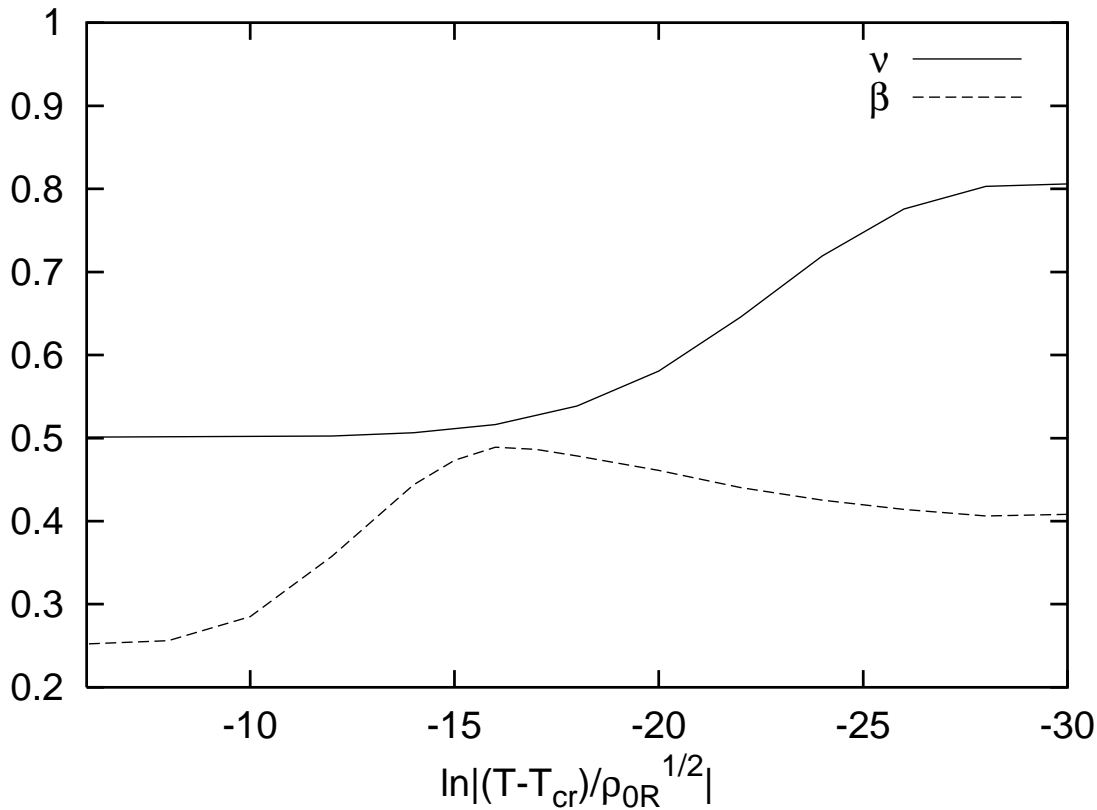,width=11cm,angle=-90}}
 \caption{\it
The critical exponents $\nu$ and $\beta$ as the critical temperature
is approached near the tricritical point.
}
 \label{fig6}
 \end{figure}

These findings are in very good agreement with the discussion of section 
2, which was based on mean field theory. The reason is that the Gaussian
fixed point results in mean-field behaviour, as the effect of 
fluctuations is negligible because of the smallness of the couplings.
Our results indicate that the tricritical point must be connected with 
the Gaussian fixed point of fig. \ref{fig5}. The number of relevant
parameters is correctly predicted to be 2. 

The effective critical exponents $\nu$, $\beta$
can be computed from eqs. (\ref{nnu}), (\ref{bbeta}), similarly to 
subsection 7.1, by keeping $\mu$ fixed and varying $T$. 
The results are depicted in fig. \ref{fig6}.
We observe that for $|T-\tcr|/\sqrt{\rho_{0R}} = \exp(-5)$ -- $\exp(-8)$
they take the values $\beta=0.25$, $\nu=0.5$. These are the mean-field
predictions that we derived in section 2. For this range of temperatures
the potential never approaches the $O(4)$ fixed point. It stays close to
the Gaussian one, before moving directly towards the broken or the symmetric
phase. For temperatures closer to $\tcr$, the
system feels the attraction of the $O(4)$ fixed point. For $T$
very close to $\tcr$ the potential 
spends a large part of its evolution near this
fixed point. As a result it loses memory of its initial form near the
Gaussian fixed point. The critical exponents take the values 
characteristic of the $O(4)$ universality class.

The curves in fig. \ref{fig6} between 
$|T-\tcr|/\sqrt{\rho_{0R}} = \exp(-9)$ and 
$|T-\tcr|/\sqrt{\rho_{0R}} = \exp(-30)$
are typical examples of crossover curves. They describe the variation of
universal quantities, such as the critical exponents, as the relative 
influence of two fixed points changes.

\begin{figure}[t]
 \centerline{\epsfig{figure=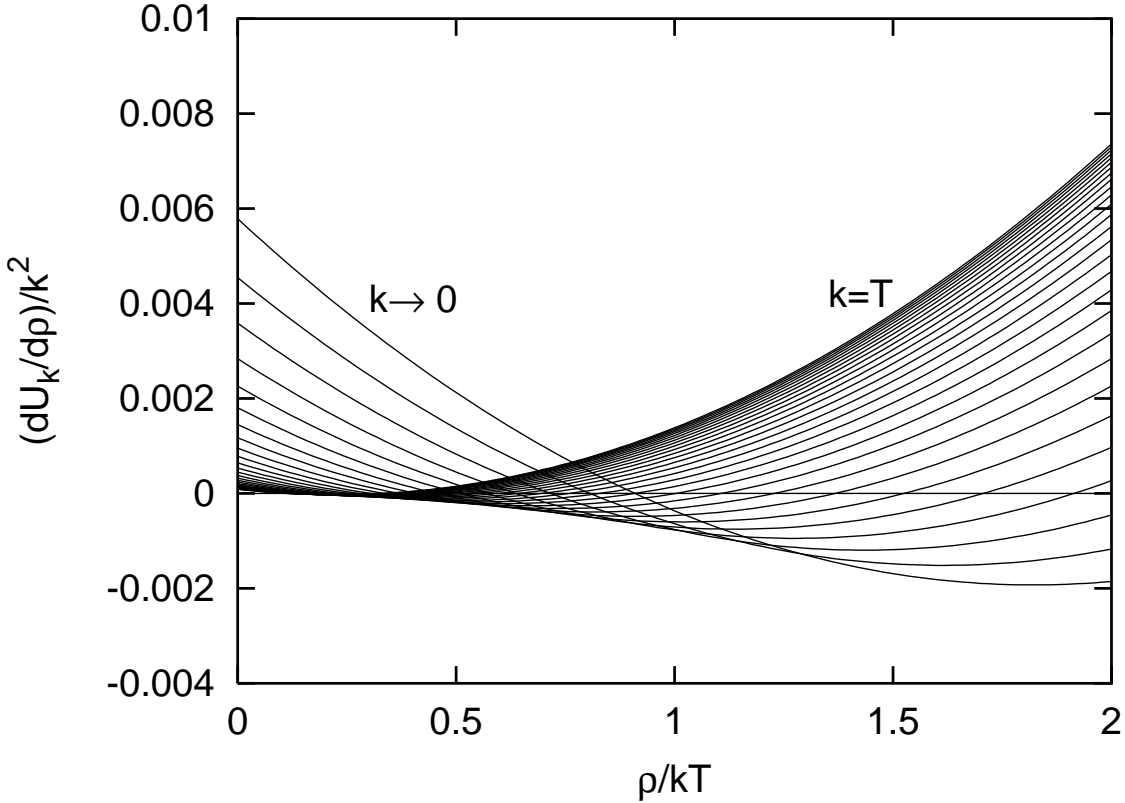,width=11cm,angle=-90}}
 \caption{\it
The evolution of the rescaled potential for 
$\mu/\sqrt{\rho_{0R}}=0.719$, $T/\sqrt{\rho_{0R}}\simeq 0.128.$
}
 \label{fig7}
 \end{figure}

We can confirm our identification of the
tricritical point by studying the first-order phase
transitions near the Gaussian fixed point.
In fig. \ref{fig7} we display the evolution of $u'_k(\rht)$ for a theory
with $\mu/\sqrt{\rho_{0R}}=0.719$, $T/\sqrt{\rho_{0R}}\simeq 0.128.$ 
The slight increase of the value of the chemical potential relative to 
fig. \ref{fig5}
modifies drastically the evolution. The potential stays very close
to the Gaussian fixed point for a long ``time'' $t=\ln(k/\rho^{1/2}_{0R})$,
but eventually moves away from it. 
(Notice the difference by a factor 100 of 
the scale of the vertical axis between figs. \ref{fig3} and \ref{fig7}.)
Its form is characteristic
of a first-order phase transition as we discussed in detail in subsection
7.2. However, the transition is very weak, as the discontinuity in the
order parameter is orders of magnitude smaller than $\rho_{0R}$.

The tricritical point is located 
on the plane ($\mu/\sqrt{\rho_{0R}}$, $T/\sqrt{\rho_{0R}}$)
in the region between
the points (0.718, 0.130) and (0.719, 0.128). 
For the critical values of $\mu$ and $\tcr$ the solution of the evolution
equation stays close to the Gaussian fixed point for an infinitely long 
``time'' $t$. As a result the critical behaviour is determined completely 
by mean-field theory. 
For example, the critical exponent $\beta$ stays equal to 0.25 arbitrarily
close to the critical temperature.

\section{The phase diagram for $j\not=0$}

\subsection{The analytical crossover and the first-order phase transitions}

The effect of a source term $-j\sx$ on a second-order phase transition
has been studied extensively in the statistical physics and field theory
literature. It is well known that the phase transition disappears and
is replaced by an analytical crossover. No singularities appear in the
physical quantities. Instead these are described by smooth  functions
connecting the two ``phases''. 
The reason can be understood by
considering the potential for a constant value of $\mu$, corresponding 
to the part of the phase diagram where a second-order
phase transition is expected for $j=0$.
The vacuum of the theory is determined by
the relation $\partial U_0/\partial \sx=j$. For $j\not=0$ the vacuum is 
located away from the origin for all values of $T$.
If we take $\sx$ as the order parameter, no signal of a phase
transition exists. Moreover, both the $\sx$-field and the pions are
massive at the vacuum. This implies that no significant fluctuations
exist in the deep infrared, 
as all fields decouple at sufficiently low energy scales. 
Consequently, no singularities are expected in physical quantities. 

For very small $j$, the influence of the fixed point leads to a universal form
of the potential in the critical region. 
It is convenient to define 
the quantity $x=\delta T/\sx^{1/\beta}$, with $\delta T=T-\tcr$ and $\beta$ a critical
exponent. The potential can be expressed as 
\be
\frac{1}{\sx^\delta} \frac{\partial U_0}{\partial \sx}=f(x),
\label{eeos} \ee
with $\delta$ another critical exponent and
$\partial U_0/\partial \sx=j$ \cite{zinn}.
The form of the function $f(x)$, characterized as the universal
equation of state, is completely specified by the 
universality class. For small $j$, 
the temperature dependence of the $\sx$-field and
pion masses 
near the critical temperature can be extracted from eq. (\ref{eeos}) 
\cite{Rajagopal:1992qz}.

We shall not determine the equation of state of the $O(4)$ universality class
here, as it has been calculated in ref. \cite{Berges:1997eu}.
There, it is also demonstrated how the interplay between universal and
non-universal behaviour can be described efficiently by the 
effective average action. Even though the study of ref. \cite{Berges:1997eu}
is carried out for $\mu=0$, it displays all the characteristic features
associated with theories in the small-$\mu$ region of the phase diagram. 

The 
first-order phase transitions for large $\mu$ that we discussed
in subsection 7.2 are not affected significantly by the presence of
a source term $-j\sx$. The minimum at the origin moves to a non-zero value
of $\sx$, as in the case of second-order phase transitions. 
The linear term also affects the relative depth of the two
minima of the potential and chages
the value of the critical temperature.
The most characteristic feature of the phase diagram for $j\not= 0$ is the 
presence of a critical endpoint that marks the end of the line of first-order
phase transitions. At this point a second-order phase transition is expected
to take place. In the following subsection we discuss in detail the
nature of the endpoint and determine the relevant universality class.

\subsection{The critical endpoint}

In order to discuss the critical endpoint 
we need to adapt our formalism to the situation of explicit symmetry
breaking through the addition of the source term $-j \sx$ to the
potential. In particular, we have to define an appropriate order parameter
that can describe the second-order phase transition.
For this reason we expand the field
$\sx$ around a non-zero value $\sxj$, so that $\sx=\sxj+\dt$ with
$\dt\ll\sxj$.
We define a new potential 
\be
V_k(\dt;T,\mu)=U_k(\sxj+\dt;T,\mu)-j\,(\sxj+\dt)
\label{newpot} \ee
and the dimensionless quantities
\begin{eqnarray}
v_k(\dtt)&=&\frac{V_k(\dt;T,\mu)}{k^3T}
\label{vkdtt} \\
\dtt&=&\frac{\dt}{\sqrt{kT}}.
\label{deltt} \end{eqnarray}
The evolution equation takes the form
\be
    \frac{\pa }{\pa t}v'_{k} (\dtt) =
      -\frac{5}{2} v'_k+\frac{1}{2}\dtt v''_k  
      +v_3 \left(v'''_k\right)
      L^3_1\left( v''_k\right)
     +3v_3\, 
      \left(
      \frac{v''_k}{\frac{\sxj}{\sqrt{kT}}+\dtt}
      -\frac{\frac{j}{\sqrt{k^5T}}+v'_k}{\left(\frac{\sxj}{\sqrt{kT}}
       +\dtt\right)^2}
      \right)
      \,L^3_1\left(\frac{\frac{j}{\sqrt{k^5T}}+v'_k}{\frac{\sxj}{\sqrt{kT}}
      +\dtt}\right),
\label{scinvj} \ee
where the primes now denote derivatives with respect to
$\dtt$.

The last term arises through the pion fluctuations. This
is apparent if one considers the argument of the threshold function, which
is the pion mass term $(1/\sx)(\pa U_k/\pa \sx)$ divided by $k^2$. The term
introduces explicit scale dependence in the equation. 
We expect
the contributions that include inverse powers of $k$ to become dominant in
the infrared. 
The potential $u'_k(\dtt)$ could also include terms that scale with
inverse powers of $k$. We isolate the most singular contribution by 
parametrizing $u'_k(\dtt)$ as
\be 
v'_k(\dtt)=\vt'_{k}(\dtt)+\frac{c_k}{\sqrt{k^5T}}.
\label{vsk} \ee
In the following we shall show that 
$\vt'_{k}(\dtt)$ can reach a scale-invariant
form during the evolution, so that $\vt'_{k}(\dtt)\ll j/\sqrt{k^5T}$ for
fixed $\dtt$ and $k\to 0$. Under this assumption, we can
write eq. (\ref{scinvj}) as 
\begin{eqnarray} 
    \frac{\pa }{\pa t}\vt'_{k} (\dtt) &=&
      -\frac{5}{2} \vt'_{k}+\frac{1}{2}\dtt \vt''_{k}  
      +v_3 \left(\vt_{k}'''\right)
      L^3_1\left( \vt''_{k}\right)
\label{lone} \\
    \frac{d }{d k}c_k  &=&
     -3v_3\, 
      \frac{(j+c_k)T}{\sxj^2}
      \,L^3_1\left(\frac{j+c_k}{\sxj k^2}\right).
\label{ltwo} 
\end{eqnarray}
These equations have a clear physical interpretation: For sufficiently
low $k$, the pions (the Goldstone modes)
decouple from the evolution of the $\sx$-field (the radial mode). The reason is
that the pions are massive in the presence of explicit chiral symmetry
breaking, with an effective mass controlled by $j$. 

The solution of eq. (\ref{lone}) can reach a scale-invariant form.
The only known fixed point of this equation
corresponds to the Ising universality class, as
only the fluctuations of the $\sx$-field contribute to the evolution.
The emergence of a critical theory around $\sx=\sxj$
imposes two requirements:\\
a) The first derivative of the potential around this point must vanish for 
$k\to 0$.\\
b) The fixed point of eq. (\ref{lone}) must be approached during the 
evolution.

The solution of eqs.(\ref{lone}), 
(\ref{ltwo}) requires initial conditions.
These are provided by the definition (\ref{newpot}) for 
$k_T=T/\theta_2$. (As before we use $\theta_2=1$ for the calculations.) 
In  particular we take
\begin{eqnarray}
c_{k_T}&=&
\frac{\pa V_{k_T}}{\pa \delta} (0;T,\mu)
=\frac{\pa U_{k_T}}{\pa \sx} (\sxj;T,\mu)-j
\label{lthree} \\
\vt'_{k_T}&=&\frac{1}{\sqrt{k_T^5T}}
\left( \frac{\pa^2 U_{k_T}}{\pa \sx^2} (\sxj;T,\mu)\, \dt
+\frac{1}{2}\frac{\pa^3 U_{k_T}}{\pa \sx^3} (\sxj;T,\mu)\, \dt^2
+\frac{1}{6}\frac{\pa^4 U_{k_T}}{\pa \sx^4} (\sxj;T,\mu)\, \dt^3
+ {\cal O}(\dt^4) \right)
\nonumber \\
&&
\label{lfour} \end{eqnarray}
and neglect the higher-order terms for small $\dt$.
The first requirement for the emergence of a critical theory
implies that $c_{k_T}$ must be
chosen such that the solution of eq. (\ref{ltwo})
gives $c_0=0$. The second requirement can be satisfied
if the 
initial condition (\ref{lfour}) has two properties: The term $\sim \dt^2$
(the cubic term) 
vanishes, and the term $\sim \dt$ (the mass term)
has the appropriate critical value for
the fixed point to be approached. The term $\sim \dt^3$ (the quartic coupling)
need not be fine-tuned,
as the Ising fixed point is characterized by one relevant parameter.
For any given value of the quartic coupling, the fine-tuning of
the mass term leads to the fixed point.
For given $j$, we have three free parameters at our disposal: $T$, $\mu$ and
the value $\sxj$ around which we expand the potential.
Therefore, we expect that 
the fixed point of eq. (\ref{lone}) can be approached for a unique
choice of $T$, $\mu$ and $\sxj$. The values of these parameters mark
the location of the critical endpoint on the phase diagram.

\begin{figure}[t]
 \centerline{\epsfig{figure=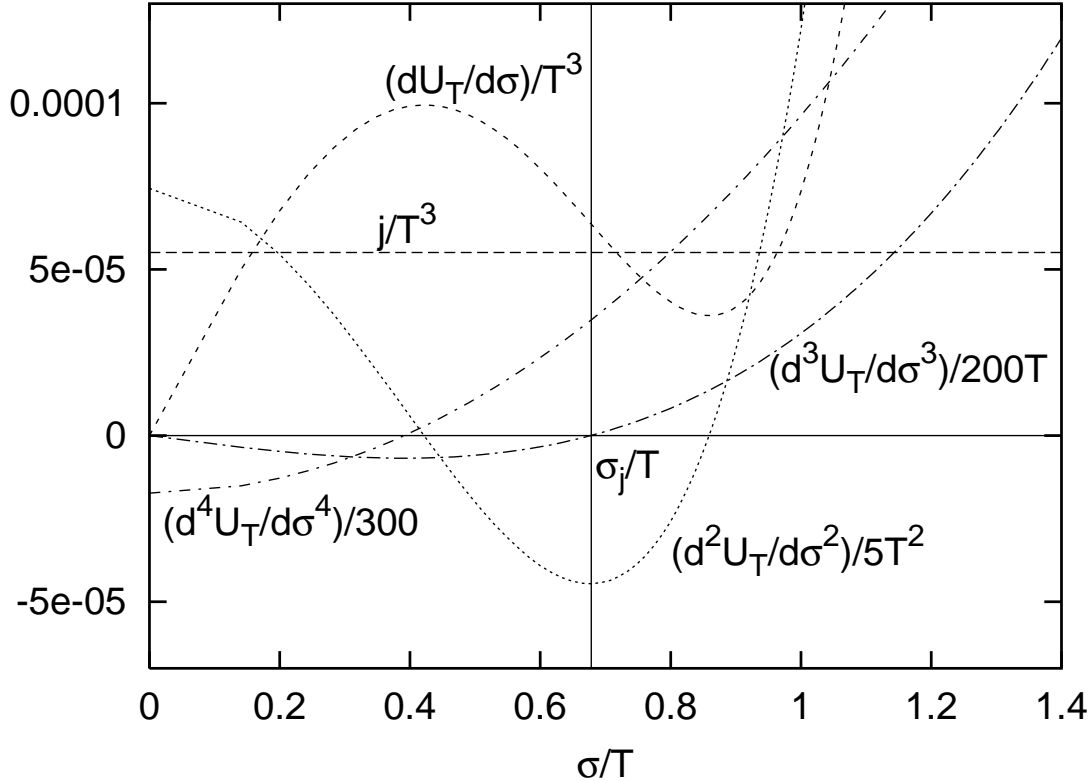,width=11cm,angle=-90}}
 \caption{\it
Determination of the initial conditions (\ref{lthree}), (\ref{lfour})
for the evolution equations
(\ref{lone}), (\ref{ltwo}).
}
 \label{fig8}
 \end{figure}

In order to determine $c_0$, eq. (\ref{ltwo}) must be solved numerically.
Alternatively, an analytical upper bound can be obtained through the relation
\be
\ln \left( 
\frac{c_0+j}{c_{k_T}+j}
\right) =  -3v_3\, 
      \frac{T}{\sxj^2}\int_{k_T}^0 dk    
      \,L^3_1\left(\frac{j+c_k}{\sxj k^2}\right).
\label{upper} \ee
The threshold function is monotonic and 
the maximum value of $|L^3_1(w)|$ is 
$|L^3_1(0)|=\sqrt{\pi}$. Thus, we obtain
\be
\left| \ln \left( 
\frac{c_0+j}{c_{k_T}+j}
\right)\right| \leq  \frac{3}{8\pi^{3/2}}
      \frac{Tk_T}{\sxj^2}. 
\label{upperbound} \ee
For our choice of parameters (see below) the r.h.s. is 
significantly smaller than 1. This implies that there is only a 
small difference between  $c_0$ and $c_{k_T}$. 
As a result, the vanishing of $c_0$ implies 
$\left[ {\pa U_{k_T}}/{\pa \sx} \right] (\sxj;T,\mu)\simeq j$, according to 
eq. (\ref{lthree}).

The determination of the initial conditions for the solution of 
eqs. (\ref{lone}), (\ref{ltwo}) is shown in fig. \ref{fig8}.
The first four derivatives of the potential $U_{k_T}$ at the initial scale 
$k_T=T$ are plotted. The point $\sx_j$ at which the third derivative
vanishes is where we expect the critical theory to emerge. 
For the parameters of the model we have chosen
($\lx=0.1$, $h^2=2.3$, $\Lx^2/\rho_{0R}=1.8$), and 
$\mu/\sqrt{\rho_{0R}}=0.72$, $T/\sqrt{\rho_{0R}}\simeq 0.127$,
we find $\sx_j/\sqrt{\rho_{0R}}\simeq 0.0858$.
We are considering an external source
$j/\rho^{3/2}_{0R}\simeq 1.12 \times 10^{-7}$. The initial value of
the parameter $c_{k_T}$, defined in eq. (\ref{lthree}), is
$c_{k_T}/\rho^{3/2}_{0R}\simeq1.74\times 10^{-8}$.
We observe that $[\partial U_{k_T}/\partial \sx](\sx_j;T,\mu)$ is close but not
exactly equal to $j$. As a result 
$[\partial V_{k_T}/\partial \delta](0;T,\mu)\not= 0$.

\begin{figure}[t]
 \centerline{\epsfig{figure=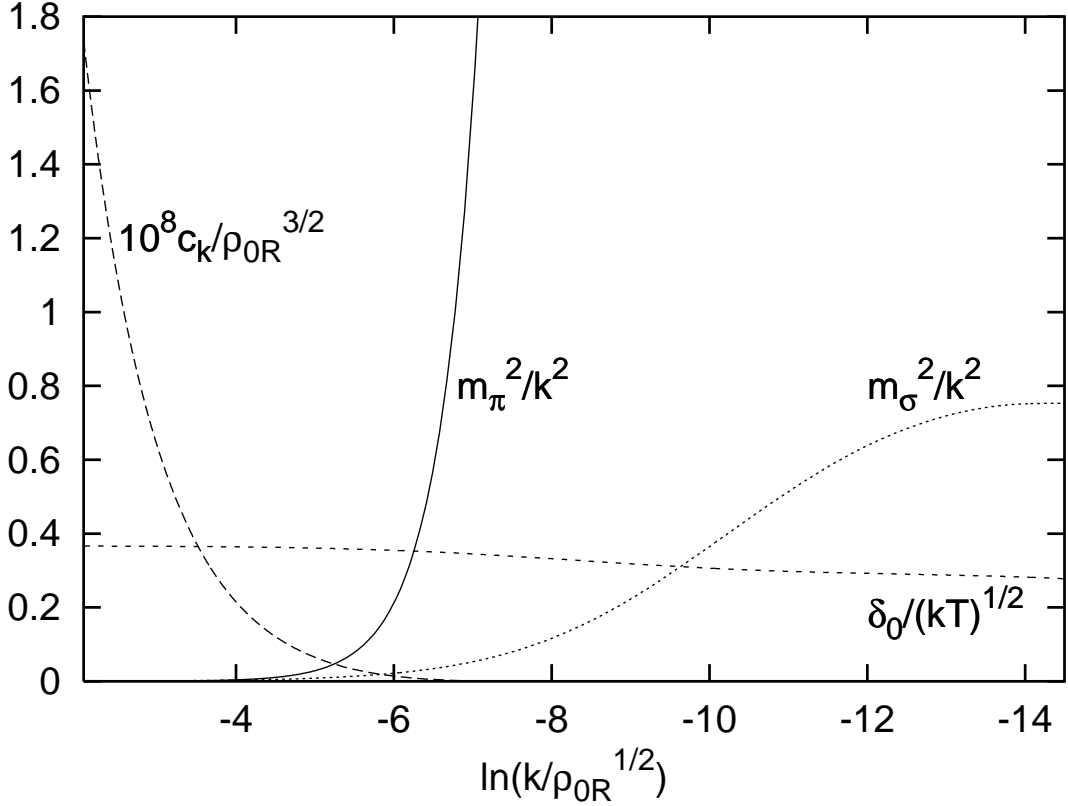,width=11cm,angle=-90}}
 \caption{\it
The evolution of: $c_k=[{\pa V_{k}}/{\pa \delta}] (0;T,\mu)$,
the rescaled $\sx$-field and pion masses, and the minimum
of $\vt_k(\dtt)=\left( V_k(\delta;T,\mu)-c_k\delta\right)/(k^3T)$.\,
$\mu/\sqrt{\rho_{0R}}=0.72$, $T/\sqrt{\rho_{0R}}\simeq 0.127$,
$j/\rho^{3/2}_{0R}\simeq 1.12 \times 10^{-7}$. 
}
 \label{fig9}
 \end{figure}

In fig. \ref{fig9} we display elements of 
the solution of eqs. (\ref{lone}), (\ref{ltwo}).
We observe that the parameter
$c_k=[{\pa V_{k}}/{\pa \delta}] (0;T,\mu)$
starts from its initial value at the scale $k_T=T$ 
and effectively becomes zero
at a scale $k/\rho_{0R}^{1/2}\simeq \exp(-6)$. The 
rescaled mass of the pion fields
$m^2_\pi(k)/k^2=[(j+c_k)/\sxj]/ k^2$
becomes much larger than 1 slightly below
this scale. As a result, the pions decouple completely from the evolution
equation and $c_k$ remains constant during the ensuing evolution.
This guarantees that the first derivative of the potential at 
$k=0$ vanishes at the point $\delta=0$.

The solution of eq. (\ref{lone}) approaches a fixed point in the infrared.
This is demonstrated in fig. \ref{fig9} by displaying: 
a) the minimum $\dtt_0=\delta_0/\sqrt{kT}$ of the rescaled potential
$\vt_k(\dtt)=\left( V_k(\delta;T,\mu)-c_k\delta\right)/(k^3T)$; and
b) the rescaled mass term 
$m^2_\sx(k)/k^2=[ \pa^2 V_k/\pa \delta^2 ](\delta_0;T,\mu)/k^2$
of the $\sx$-field (or the shifted $\delta$-field) at the minimum.
Both these quantities reach constant values at a scale
$k/\rho_{0R}^{1/2}\simeq \exp(-14)$. The subsequent 
evolution is analogous to the one studied in subsection 7.1. The
potential, and the various parameters derived from it, stay
constant near their fixed-point values for some ``time'' 
$t=\ln(k/\rho^{1/2}_{0R})$,
before finally deviating towards the
symmetric or the broken phase. The renormalized theory at $k=0$ exhibits
a second-order phase
transition. The parameter that has to be fine-tuned in order to
go through the critical endpoint is a linear combination of $T$ and
$\mu$.

The universality class of the fixed point is determined by the 
number of fields with running masses comparable or smaller than $k$ 
and the symmetries of the relevant part of the action. 
For $k/\rho_{0R}^{1/2}\lta \exp(-6)$ only the 
$\sx$-field fulfills this criterion. The pions have completely decoupled by
the time the fixed point is approached. 
The scale-dependent potential $\vt(\dtt)$ has a $Z_2$ symmetry  
that is preserved by the evolution. Crucial for this conclusion is the
choice of a point $\sx_j$ at which the cubic term of the potential 
$U_{k_T}$ vanishes. The elimination of the pion contribution from
the evolution equation (\ref{lone}) through the parametrization 
(\ref{vsk}) is important as well. The pion fluctuations only affect
the evolution of the term linear in $\delta$ around $\sx_j$ (the parameter
$c_k$).  
This term evolves to zero and stays constant after the pion decoupling. 
These features can be achieved for specific values of $T$ and $\mu$ 
that determine the location of the critical endpoint.
As a result, the endpoint
belongs to the Ising universality class. 
This can be confirmed by calculating the critical exponents, as
was done in subsection 7.1.

Before concluding this section, we should comment on the
approximations employed in the study of the critical endpoint.
The parametrization of eq. (\ref{vsk}) is quite general. 
The derivation of the evolution equations (\ref{lone}), (\ref{ltwo}) made
use of the crucial assumption 
$\vt'_{k}(\dtt)\ll c_k/\sqrt{k^5T}$. In the example we presented, the
starting scale $k_T=T$ was sufficiently low for this relation to 
be satisfied to a good approximation during the whole evolution. 
We also neglected the terms 
higher than $\delta^3$ in the initial condition (\ref{lfour})
for $\vt'_{k_T}$. Again, this is a reasonable approximation if we are
interested only in a small region around $\sx_j$. 

Going beyond these approximations in order to increase the 
precision of the study leads to significant technical complications. 
The reason is that the $Z_2$ symmetry of $\vt_k(\dtt)$ is not guaranteed
automatically if the approximations are relaxed.
However, we believe that the essential conclusions remain
valid. In order to have a critical theory at a point $\sx_j$ one
must ensure that, at $k=0$, the first and third derivatives of the 
potential are zero there. These two conditions guarantee that $\sx_j$ would
be a minimum, 
and that no second minimum would exist in its vicinity. The appropriate 
choice of $\sx_j$ and the fine-tuning of a linear combination of
$\mu$ and $T$ can achieve this goal. The fine-tuning of the orthogonal
combination of $\mu$ and $T$ can be used in order to approach the 
fixed point that controls the critical theory. As the Ising fixed point is
the only one known for the one-field theory 
after the pion decoupling, the resulting critical
theory must be in the Ising universality class.

The demonstration that this scenario is realized, if no approximations
for the potential are made, must confront 
non-trivial technical complications. It is difficult to keep track 
simultaneously of the ``non-universal'' part of the potential corresponding
to its first and third derivatives, and the ``universal'' part that 
approaches the fixed point. The reason is that the two parts scale differently
with $k$.
Our approximations in this work simplified this
task by separating the evolution of $c_k$ from that of $\vt'(\dtt)$. 
The discussion of the general case merits a separate study.

\section{Summary and outlook}

The purpose of this study was to establish the appropriate framework for
the analytical study of the QCD phase diagram, for the 
relatively low values of the chemical potential $\mu$ relevant for
heavy-ion experiments. We employed an effective
model of QCD, the linear quark-meson model. We considered only the case of
two flavours, as all the expected structure of the phase diagram appears
already at this level. The model is 
similar to the $\sigma$-model of Gell-Mann and Levy
\cite{Gell-Mann:np}, with
the two nucleons replaced by the $u$ and $d$ quarks. However, it can be
extended in a straightforward manner in order to take into account more
flavours \cite{Jungnickel:1995fp}.

An efficient framework must
involve the renormalization group, in order to provide a reliable 
identification of the universality classes of the various second-order
phase transitions. The Wilsonian (exact) 
formulation of the renormalization group permits also the detailed description
of weak first-order phase transitions. 
We employed the formalism of the effective average action
\cite{Wetterich:1989xg}. 
We concentrated on the evolution of 
the non-derivative part of the action, the mesonic potential $U_k$, 
as an effective infrared cutoff $k$ 
is varied. The dependence of the potential on $k$ is described by an
evolution equation that takes the form of a
partial differential equation. 
The potential at a scale $k_T$ equal to the tempeature $T$
serves as the initial condition.
For $k=0$ the solution becomes equal to the effective potential, from
which all physical information relevant for the vacuum structure can
be extracted. 

We considered values of the parameters of the model that do not
correspond to realistic QCD. However, they were chosen such that the
discussion could be carried out with the minimum use of numerics\footnote{
The parameters were also chosen such that the pattern of
a second-order phase transition for $\mu=j=0$ and increasing $T$ and
a first-order one for $T=j=0$ and increasing $\mu$ appears.
The realistic values have been shown to predict this pattern 
\cite{Berges:1998sd,Berges:1997eu}.
}. 
In particular, the potential at $k_T=T$ was 
derived analytically, using perturbation theory. Its form is
determined by the potential at zero $T$ and $\mu$ and the contributions
from fluctuations with characteristic momenta larger than $k_T$.
The fermionic fluctuations play a crucial role, as they can change
the vacuum strucure of the potential by generating additional minima. 
They decouple at scales below $T$, because of the 
effective mass developed by the fermions at non-zero temperature.
As a result, the entire fermionic contributions and the mesonic
contributions with momenta larger than $k_T$ can be incorporated
in the potential at $k_T$.
The evolution below $k_T$ is effectively three-dimensional
(dimensional reduction), as only the
zero-modes of the mesonic fields are light at these scales. 

The other significant simplification afforded by our choice of couplings
was the possibility to neglect the wavefunction renormalization of both
the fermionic and mesonic fields, and the evolution of the 
Yukawa coupling of the meson-fermion interaction. 
This meant that we could use a very simple
truncation of the effective average action, which involves standard kinetic
terms and a general scale-dependent potential for the $\sx$ and
pion fields of the mesonic sector.
The evolution of the potential was determined by solving numerically
a partial differential equation. As we mentioned above, the fermionic
fluctuations affect only the initial condition for this equation.

The phase diagram we derived has all the expected features. 
For zero current quark masses, there is a line of second-order 
phase transitions starting on the $\mu=0$ axis. We confirmed that
they belong to the
$O(4)$ universality class, by calculating universal quantities
such as the critical exponents $\beta$ and $\nu$. For large $\mu$ there is a
line of first-order phase transitions. 

The two lines meet at a tricritical
point, with specific values ($\mu_*$, $T_*$),
where a second-order phase transition takes place, governed
by the Gaussian fixed point. This results in mean-field behaviour, as 
we checked by calculating the relevant critical exponents.
For $\mu$ slightly smaller than $\mu_*$ we observed universal 
crossover behaviour
(not to be confused with the analytical crossover), as
the initial influence of the Gaussian fixed point is slowly dominated
by the more stable $O(4)$ fixed point near the critical temperature.
For $\mu$ slightly larger than $\mu_*$ we observed very weak first-order
phase transitions, for which the discontinuity in the order parameter
approaches zero.

We then took into account the effects of non-zero current quark masses
by introducing a source term $-j\sx$. We discussed how the second-order
phase transitions turn into an analytical crossover, while the first-order
ones retain their qualitative character. We studied in detail the
emergence of the critical endpoint of the line of first-order phase
transitions. The pions are massive near the endpoint and we demonstrated
their decoupling from the evolution equation. 
A second-order phase transition takes place at the endpoint, and we showed
that it belongs to the Ising universality class. 

The study demonstrated that it is possible to obtain a complete and detailed
picture of the QCD phase diagram using an effective theory such as the
quark-meson model. The renormalization group is indispensable in the process.

Realistic values for parameters of
the model cannot be considered without having to rely
on extensive numerical work. The most important issue concerns the 
value of the Yukawa coupling, which must significantly larger than 1 
for the constituent quark masses to be reproduced
correctly \cite{Jungnickel:1996fd}. 
This implies that the evolution of this coupling
and the wavefunction renormalization of the fields must be taken into account.
Of particular interest is the emergence of partial infrared fixed points 
at zero temperature and chemical potential 
\cite{Jungnickel:1996fd,Berges:1997eu}.
The derivation of the initial conditions for the effectively three-dimensional
evolution at scales below $T$ cannot be achieved analytically for large Yukawa
couplings. However, it can be performed efficiently through the
numericall integration of the appropriate evolution equations 
\cite{Berges:1997eu}.

One technical issue related to the dimensional reduction of the theory at
energy scales below $T$ concerns the fermionic infrared cutoff
that is used in the definition of the effective average action.
A form must be devised that 
preserves the Lorentz structure of the kinetic term of free fermions
for non-zero chemical potential,
and in the same time guarantees fast
decoupling of the temperature effects at low energy scales.

Finally, the strange quark can be taken into account by considering
the linear quark-meson model 
with an $SU(3)_L \times SU(3)_R$ chiral flavour group. The increase
in complexity of the
resulting evolution equations is significant. However, the process can
be carried out in a straightforward manner along the lines of refs. 
\cite{Jungnickel:1995fp}.

The present study has established that the linear quark-meson model is
an effective theory of QCD that predicts all the expected qualitative structure
of the phase diagram.
The main aim of future studies along the lines highlighted above
will be to determine with accuracy the location of the
critical endpoint and the size of the area around it in which 
the critical behaviour persists. In this way, the possible 
effect of the endpoint
on observable signatures in heavy-ion experiments
will be determined quantitatively.


\newpage



\begin{thebibliography}{999} 

\bibitem{Rajagopal:2000wf}
K.~Rajagopal and F.~Wilczek,
arXiv:hep-ph/0011333.

\bibitem{Asakawa:bq}
M.~Asakawa and K.~Yazaki,
Nucl.\ Phys.\ A {\bf 504} (1989) 668.

\bibitem{Barducci:1993bh}
A.~Barducci, R.~Casalbuoni, S.~De Curtis, R.~Gatto and G.~Pettini,
Phys.\ Lett.\ B {\bf 231} (1989) 463;
Phys.\ Rev.\ D {\bf 41} (1990) 1610;
Phys.\ Rev.\ D {\bf 42} (1990) 1757;
A.~Barducci, R.~Casalbuoni, G.~Pettini and R.~Gatto,
Phys.\ Rev.\ D {\bf 49} (1994) 426.

\bibitem{Klevansky:qe}
S.~P.~Klevansky,
Rev.\ Mod.\ Phys.\  {\bf 64} (1992) 649.

\bibitem{Stephanov:1996ki}
M.~A.~Stephanov,
Phys.\ Rev.\ Lett.\  {\bf 76}, 4472 (1996)
[arXiv:hep-lat/9604003].

\bibitem{Alford:1997zt}
M.~G.~Alford, K.~Rajagopal and F.~Wilczek,
Phys.\ Lett.\ B {\bf 422} (1998) 247
[arXiv:hep-ph/9711395].

\bibitem{Halasz:1998qr}
M.~A.~Halasz, A.~D.~Jackson, R.~E.~Shrock, M.~A.~Stephanov and J.~J.~Verbaarschot,
Phys.\ Rev.\ D {\bf 58} (1998) 096007
[arXiv:hep-ph/9804290].

\bibitem{Berges:1998rc}
J.~Berges and K.~Rajagopal,
Nucl.\ Phys.\ B {\bf 538} (1999) 215
[arXiv:hep-ph/9804233].

\bibitem{Harada:1998zq}
M.~Harada and A.~Shibata,
Phys.\ Rev.\ D {\bf 59} (1999) 014010
[arXiv:hep-ph/9807408].

\bibitem{Kiriyama:2000yp}
O.~Kiriyama, M.~Maruyama and F.~Takagi,
Phys.\ Rev.\ D {\bf 62} (2000) 105008
[arXiv:hep-ph/0001108];
Phys.\ Rev.\ D {\bf 63} (2001) 116009
[arXiv:hep-ph/0101110].

\bibitem{Fodor:2001pe}
Z.~Fodor and S.~D.~Katz,
Phys.\ Lett.\ B {\bf 534} (2002) 87
[arXiv:hep-lat/0104001];
JHEP {\bf 0203} (2002) 014
[arXiv:hep-lat/0106002].

\bibitem{deForcrand:2002ci}
P.~de Forcrand and O.~Philipsen,
Nucl.\ Phys.\ B {\bf 642} (2002) 290
[arXiv:hep-lat/0205016].

\bibitem{Allton:2002zi}
C.~R.~Allton {\it et al.},
Phys.\ Rev.\ D {\bf 66} (2002) 074507
[arXiv:hep-lat/0204010].

\bibitem{Wilson:1973jj}
K.~G.~Wilson and J.~B.~Kogut,
Phys.\ Rept.\  {\bf 12} (1974) 75.

\bibitem{rome}
See, for example, the proceedings of the 
{\it 2nd Conference on the Exact Renormalization Group}, 
Rome, Italy, 18-22 Sep 2000, and references therein. 

\bibitem{Wetterich:1989xg}
C.~Wetterich,
Nucl.\ Phys.\ B {\bf 352} (1991) 529;
C.~Wetterich,
Z.\ Phys.\ C {\bf 57} (1993) 451.

\bibitem{Wetterich:yh}
C.~Wetterich,
Phys.\ Lett.\ B {\bf 301} (1993) 90.

\bibitem{Tetradis:1993ts}
N.~Tetradis and C.~Wetterich,
Nucl.\ Phys.\ B {\bf 422} (1994) 541
[arXiv:hep-ph/9308214].

\bibitem{Berges:2000ew}
J.~Berges, N.~Tetradis and C.~Wetterich,
Phys.\ Rept.\  {\bf 363} (2002) 223
[arXiv:hep-ph/0005122].

\bibitem{Reuter:1993kw}
M.~Reuter and C.~Wetterich,
Nucl.\ Phys.\ B {\bf 417} (1994) 181;
Nucl.\ Phys.\ B {\bf 427} (1994) 291.

\bibitem{Ellwanger:iz}
U.~Ellwanger,
Phys.\ Lett.\ B {\bf 335} (1994) 364
[arXiv:hep-th/9402077];
U.~Ellwanger, M.~Hirsch and A.~Weber,
Z.\ Phys.\ C {\bf 69} (1996) 687
[arXiv:hep-th/9506019].

\bibitem{Bonini:1994kp}
M.~Bonini, M.~D'Attanasio and G.~Marchesini,
Nucl.\ Phys.\ B {\bf 437} (1995) 163
[arXiv:hep-th/9410138];
Phys.\ Lett.\ B {\bf 346} (1995) 87
[arXiv:hep-th/9412195].

\bibitem{Morris:1999px}
T.~R.~Morris,
Nucl.\ Phys.\ B {\bf 573} (2000) 97
[arXiv:hep-th/9910058];
JHEP {\bf 0012} (2000) 012
[arXiv:hep-th/0006064];
S.~Arnone, A.~Gatti and T.~R.~Morris,
arXiv:hep-th/0209162.

\bibitem{Gell-Mann:np}
M.~Gell-Mann and M.~Levy,
Nuovo Cim.\  {\bf 16} (1960) 705.

\bibitem{Jungnickel:1995fp}
D.~U.~Jungnickel and C.~Wetterich,
Phys.\ Rev.\ D {\bf 53} (1996) 5142
[arXiv:hep-ph/9505267];
Eur.\ Phys.\ J.\ C {\bf 1} (1998) 669
[arXiv:hep-ph/9606483];
Phys.\ Lett.\ B {\bf 389} (1996) 600
[arXiv:hep-ph/9607411].

\bibitem{Berges:1998sd}
J.~Berges, D.~U.~Jungnickel and C.~Wetterich,
Eur.\ Phys.\ J.\ C {\bf 13} (2000) 323
[arXiv:hep-ph/9811347].

\bibitem{Berges:1997eu}
J.~Berges, D.~U.~Jungnickel and C.~Wetterich,
Phys.\ Rev.\ D {\bf 59} (1999) 034010
[arXiv:hep-ph/9705474].

\bibitem{Ellwanger:wy}
U.~Ellwanger and C.~Wetterich,
Nucl.\ Phys.\ B {\bf 423} (1994) 137
[arXiv:hep-ph/9402221].

\bibitem{Jungnickel:1996fd}
D.~U.~Jungnickel and C.~Wetterich,
arXiv:hep-ph/9610336.

\bibitem{Nambu:tp}
Y.~Nambu and G.~Jona-Lasinio,
Phys.\ Rev.\  {\bf 122} (1961) 345.

\bibitem{Tetradis:1992qt}
N.~Tetradis and C.~Wetterich,
Nucl.\ Phys.\ B {\bf 383} (1992) 197.

\bibitem{Tetradis:1992xd}
N.~Tetradis and C.~Wetterich,
Nucl.\ Phys.\ B {\bf 398} (1993) 659.

\bibitem{Tetradis:1996fw}
N.~Tetradis,
Nucl.\ Phys.\ B {\bf 488} (1997) 92
[arXiv:hep-ph/9608272].

\bibitem{Dolan:qd}
L.~Dolan and R.~Jackiw,
Phys.\ Rev.\ D {\bf 9} (1974) 3320.




\bibitem{Strumia:1998nf}
A.~Strumia and N.~Tetradis,
Nucl.\ Phys.\ B {\bf 542} (1999) 719
[arXiv:hep-ph/9806453];
Nucl.\ Phys.\ B {\bf 554} (1999) 697
[arXiv:hep-ph/9811438];
Nucl.\ Phys.\ B {\bf 560} (1999) 482
[arXiv:hep-ph/9904246];
JHEP {\bf 9911} (1999) 023
[arXiv:hep-ph/9904357];
A.~Strumia, N.~Tetradis and C.~Wetterich,
Phys.\ Lett.\ B {\bf 467} (1999) 279
[arXiv:hep-ph/9808263].

\bibitem{zinn}
J. Zinn-Justin, {\it Quantum Field Theory and Critical Phenomena},
Oxford University Press, Oxford, 1989.

\bibitem{Rajagopal:1992qz}
K.~Rajagopal and F.~Wilczek,
Nucl.\ Phys.\ B {\bf 399} (1993) 395
[arXiv:hep-ph/9210253].







\end{thebibliography}
\end{document}